\documentclass{aa}  
\usepackage{soul}
\usepackage{graphicx}
\usepackage{multicol}
\usepackage{float}
\usepackage{caption}
\usepackage{multirow}
\usepackage{enumitem}
\usepackage{tablefootnote}
\usepackage{amssymb}
\usepackage{pifont}
\usepackage{natbib}
\usepackage{url}
\usepackage{txfonts}
\usepackage{hyperref}
\usepackage{placeins}
\usepackage{orcidlink}
\usepackage{xcolor}
\defcitealias{XRISM_2025}{Paper I}
\defcitealias{Simionescu_2026}{Paper II}
\bibpunct{(}{)}{;}{a}{}{,}

\begin{document}

    \title{High-resolution X-ray spectroscopy with XRISM/Resolve reveals super-Solar abundance ratios in Virgo/M87}
    \titlerunning{Super-Solar abundance ratios in M87}
    \authorrunning{J. Martin}
    \titlerunning{Super-Solar abundance ratios in M87}
    \author{J. Martin \inst{1,2}\orcidlink{0009-0007-5164-8121}
    \and A. Simionescu \inst{1,2,3}\orcidlink{0000-0002-9714-3862}
    \and F. Mernier \inst{18,9,4,10} \orcidlink{0000-0002-7031-4772}
    \and C. Kilbourne \inst{4} \orcidlink{0000-0001-9464-4103}
    \and A. T\"umer \inst{12,4,10} \orcidlink{0000-0002-3132-8776}
    \and H.~R. Russell \inst{5} \orcidlink{0000-0001-5208-649X}
    \and M. Charbonneau \inst{7} 
    \and N. Dizdar \inst{7} \orcidlink{0009-0003-9080-6736}
    \and D. Eckert \inst{8} \orcidlink{0000-0001-7917-3892}
    \and Y. Ezoe \inst{13}
    \and R. Fujimoto \inst{14} \orcidlink{0000-0002-2374-7073}
    \and M. Fujita \inst{15}
    \and K. Fukushima \inst{22} \orcidlink{0000-0001-8055-7113}
    \and L. Gu \inst{1} \orcidlink{0000-0001-9911-7038}
    \and E. Hodges-Kluck \inst{4} \orcidlink{0000-0002-2397-206X}
    \and Y. Ichinohe \inst{15} \orcidlink{0000-0002-6102-1441}
    \and D. Ito \inst{6} \orcidlink{0009-0000-4742-5098}
    \and S. Kitamoto \inst{16} \orcidlink{0000-0001-8948-7983}
    \and M. A. Leutenegger \inst{4} \orcidlink{0000-0002-3331-7595}
    \and M. Loewenstein \inst{9,4,10} \orcidlink{0000-0002-1661-4029}
    \and H. McCall \inst{11} \orcidlink{0000-0003-3537-3491}
    \and B.~R. McNamara \inst{7}
    \and E. D. Miller \inst{19} \orcidlink{0000-0002-3031-2326}
    \and I. Mitsuishi \inst{6} \orcidlink{0000-0002-9901-233X}
    \and K. Nakazawa \inst{6} \orcidlink{0000-0003-2930-350X}
    \and A. Ogorzalek \inst{9,4,10} \orcidlink{0000-0003-4504-2557}
    \and K. Sato \inst{20} \orcidlink{0000-0001-5774-1633}
    \and A. Szymkowiak \inst{21} \orcidlink{0000-0002-4974-687X}
    \and I. Zhuravleva \inst{11} \orcidlink{0000-0001-7630-8085}    
    }
    \institute{
        SRON, Space Research Organisation Netherlands, Niels Bohrweg 4, 2333 CA Leiden, The Netherlands 
        \and Leiden Observatory, Leiden University, PO Box 9513, 2300 RA Leiden, The Netherlands
        \and Kavli Institute for the Physics and Mathematics of the Universe (WPI), The University of Tokyo, Kashiwa, Chiba 277-8583, Japan
        \and NASA / Goddard Space Flight Center, Greenbelt, MD 20771, USA
        \and School of Physics \& Astronomy, University of Nottingham, Nottingham, NG7 2RD, UK
        \and Department of Physics, Nagoya University, Aichi 464-8602, Japan
        \and Department of Physics \& Astronomy, Waterloo Centre for Astrophysics, University of Waterloo, Ontario N2L 3G1, Canada
        \and Department of Astronomy, University of Geneva, Versoix CH-1290, Switzerland
        \and Department of Astronomy, University of Maryland, College Park, MD 20742, USA
        \and Center for Research and Exploration in Space Science and Technology, NASA / GSFC (CRESST II), Greenbelt, MD 20771, USA
        \and Department of Astronomy and Astrophysics, University of Chicago, Chicago, IL 60637, USA
        \and Center for Space Sciences and Technology, University of Maryland, Baltimore County (UMBC), Baltimore, MD, 21250 USA
        \and Department of Physics, Tokyo Metropolitan University, Tokyo 192-0397, Japan
        \and Institute of Space and Astronautical Science (ISAS), Japan Aerospace Exploration Agency (JAXA), Kanagawa 252-5210, Japan
        \and Department of Physics, Saitama University, Saitama 338-8570, Japan
        \and RIKEN Nishina Center, Saitama 351-0198, Japan
        \and Department of Physics, Rikkyo University, Tokyo 171-8501, Japan
        \and Univ Toulouse, CNES, CNRS, IRAP, Toulouse, France
        \and Kavli Institute for Astrophysics and Space Research, Massachusetts Institute of Technology, MA 02139, USA
        \and Department of Astrophysics and Atmospheric Sciences, Kyoto Sangyo University, Kyoto 603-8555, Japan
        \and Yale Center for Astronomy and Astrophysics, Yale University, CT 06520-8121, USA
        \and Department of Physics, Tokyo University of Science, 1-3 Kagurazaka, Shinjuku-ku, Tokyo 162-8601, Japan
        }
   \date{Received XXX / Accepted XXX}
  \abstract{The chemical composition of the intracluster medium (ICM) provides key insights into the enrichment history of galaxy clusters. However, high-resolution abundance measurements with X-ray microcalorimeters remain available for only a few systems. While most cool-core clusters exhibit near-Solar elemental abundance ratios relative to Fe, previous studies of the Virgo cluster have suggested super-Solar ratios in its core.}{We investigate the chemical properties of the Virgo cluster core using \textit{XRISM}/Resolve observations, focusing on precise measurements of Si, S, Ar, Ca, Cr, Fe, and Ni abundances. We aim to determine whether Virgo displays abundance patterns distinct from other nearby cool-core clusters and to explore the physical origin of any differences.}{We analyzed full-field-of-view \textit{XRISM}/Resolve spectra in four regions of the Virgo core (center, east, northwest, and southwest) in the 1.7--11 keV band. Single-temperature, multi-temperature, and multi-abundance models were applied to characterize the thermal structure and derive elemental abundances. The resulting abundance ratios were compared between pointings, with previous studies of Virgo/M87, and with recent \textit{XRISM} measurements of other clusters.}{All four pointings exhibit systematically super-Solar X/Fe ratios, although the northwest region shows values closer to Solar. Multi-abundance modeling of the eastern and southwestern regions reveals that cool, metal-rich gas uplifted by the AGN coexists with a hotter, more chemically homogeneous ambient ICM. The super-Solar ratios are robust against variations in bandpass and temperature structure. We interpret these enhanced ratios as reflecting the enrichment history of the old stellar population in M87 combined with a limited cold gas reservoir.}{The Virgo core exhibits spatially resolved chemical enrichment that differs from the abundance patterns observed in other cool-core clusters. These results highlight the importance of high-resolution spectroscopy for disentangling AGN-driven transport, multiphase structure, and stellar enrichment. Expanding microcalorimeter observations to a larger cluster sample will be essential to determine whether Virgo is an outlier or representative of a broader diversity in cluster-core enrichment histories.}
\maketitle
\section{Introduction}

Galaxy clusters are excellent laboratories for studying the chemical enrichment history of the Universe. They retain the metals produced in their member galaxies within the intracluster medium (ICM), preserving an integrated record over cosmic time of early pre-enrichment, star formation, stellar activity, galaxy interaction and Active Galactic Nuclei (AGN) feedback. Indeed, AGN are invoked to transport heavy elements into the ICM \citep[e.g. ][]{Rebusco_2006, Roediger_2007} which contains the majority of the baryonic mass of clusters. The ICM is a hot ($10^{7}$–$10^{8}$ K) and diffuse plasma that emits primarily in X-rays through thermal bremsstrahlung and line emission from highly ionized metals (K- and L-shell transitions). X-ray spectroscopy thus provides an ideal probe of elemental abundances. To date, X-ray observations of the ICM have enabled measurements of the abundances of Fe, Ni, and other Fe-peak elements, as well as $\alpha$-elements such as O, Ne, Mg, Si, S, Ar, and Ca. Fe-group/Fe ratios trace Type Ia supernovae, while $\alpha$/Fe ratios trace core-collapse (CC) to Type Ia supernovae (SNe) contributions and enrichment timescales, and lighter elements probe AGB stars and the initial mass function.

The first detections of metal line emission in the ICM were achieved with Ariel V in Perseus \citep{Mitchell_1976} and with OSO-8 \citep{Serlemitsos_1977}. Subsequent observations with ROSAT and ASCA enabled the first measurements of abundance ratios, revealing approximately Solar $\alpha$/Fe ratios in many systems and central metallicity enhancements (e.g. \citealt{Mushotzky_1996}, \citealt{Baumgartner_2005}). A major improvement followed with the Chandra X-ray Observatory, XMM-Newton, and Suzaku, whose increased effective area and improved spectral capabilities provided more precise constraints on elements from O to Ni, and in deep exposures Cr and N. These studies generally reported near-Solar X/Fe ratios and radial variations (e.g. \citealt{Tamura_2001}; \citealt{Werner_2006}; \citealt{De_Plaa_2007}; \citealt{Mernier_2016a}).

Although abundance ratios reported in the literature are generally consistent with Solar values, some measurements span a range from slightly sub-Solar to mildly super-Solar \citep{Matsushita_2007, De_Plaa_2017, Mernier_2016a, XRISM2025_Centaurus}, depending on the element, source, and analysis assumptions. Moreover, the limited spectral resolution of CCD detectors leads to line blending, and uncertainties in atomic modeling affect the robustness of X/Fe measurements. High-resolution X-ray spectroscopy provides the most direct and reliable means to disentangle blended emission lines and, along with robust spectral modeling and updated atomic databases, accurately determine elemental abundances in the ICM, making it an essential tool for reconstructing the chemical enrichment history of galaxy clusters (for a review see \citealt{Mernier_2018_review}). 
 
The short-lived \textit{Hitomi} mission demonstrated the potential of non-dispersive, high-resolution spectroscopy. Its observations of the Perseus cluster core revealed abundance ratios consistent with Solar values \citep{Hitomi_2017,Simionescu_2019}. The launch of the X-Ray Imaging and Spectroscopy Mission (\textit{XRISM}) \citep{Tashiro_2025} restores and extends these capabilities with the Resolve microcalorimeter \citep{Kelley_2025,Ishisaki_2025}, enabling clean separation of key transitions from Si, S, Ar, Ca, Cr, Fe, and Ni, and significantly reducing systematic uncertainties compared to CCD-based measurements \citep{Hitomi_2017, XRISM2025_Centaurus, XRISM2025_Coma, Sarkar_2025}. High-resolution spectroscopy is therefore crucial for obtaining robust and less biased measurements of the ICM composition, by mitigating the limitations inherent to CCD-based analysis.

It remains unclear whether the chemical composition of cluster cores is universal, or instead shaped by local enrichment history, multiphase structure, and AGN-driven metal transport. In this context, Virgo, with its AGN-hosting and brightest cluster galaxy (BCG) M87, provides an ideal target. At a distance of 16.4 Mpc \citep{Rackers_2024}, it is the second nearest extragalactic X-ray bright source, offering unmatched spatial leverage and high surface brightness. Deep observations with Chandra and XMM-Newton have revealed a dynamically complex, multiphase ICM shaped by AGN feedback, including cavities, shocks, and uplifted low-entropy gas along X-ray bright arms that are believed to be driven by the AGN jet and counter-jet \citep[e.g.][]{Feigelson_1987,Bohringer_1995,Belsole_2001, Young_2002,Forman_2005,Werner_2010,Million_2010}. M87 is surrounded by an extensive gaseous atmosphere \citep[e.g.][]{McCall_2024} with temperatures averaging around $\sim1-3$~keV \citep[e.g.][]{Matsushita_2002, Simionescu_2008}, ensuring an X-ray spectrum rich in metal emission lines that allow us to perform detailed diagnostics of the chemical abundances. Previous abundance studies based primarily on CCD-resolution spectroscopy have reported centrally peaked metallicity profiles and significant radial variations. While some studies found sub-Solar \citep[e.g.][]{Finoguenov_2000} or near-Solar $\alpha$/Fe \citep[e.g.][]{Mernier_2016a}, others found hints of super-Solar ratios \citep{Gatuzz_2023}. In the Virgo core, the presence of multiple temperature components \citep{Belsole_2001,Matsushita_2002,Molendi_2002,Simionescu_2008}, projection effects, and point-spread function scattering between regions can bias abundance measurements if not properly accounted for. High-resolution spectroscopy with \textit{XRISM}/Resolve is therefore crucial to separate temperature phases and obtain reliable X/Fe ratios in this dynamically complex environment. 

In this paper, we present a detailed analysis of the elemental abundances in the ICM surrounding M87 in the Virgo cluster, using high-resolution X-ray spectra obtained with \textit{XRISM}/Resolve. By exploiting the instrument's spectral resolving power, we are able to model emission lines with unprecedented precision and to assess the robustness of abundance measurements across four spatial regions. In Sect.~\ref{sec:Observations and Data Reduction}, we describe the observations and data reduction procedures; Sect.~\ref{sec:Data analysis} includes the spectral modeling approaches; Sect.~\ref{sec:Results} presents the results, focusing on derived abundance ratios, and an assessment of systematic uncertainties; in Sect.~\ref{sec:Discussion}, we discuss the implications of our findings in the context of previous studies of Virgo and recent \textit{XRISM} studies of other clusters. Sect.~\ref{sec:Conclusion} summarizes our conclusions and outlines prospects for future \textit{XRISM} observations of galaxy clusters.

Throughout this paper, we adopt the cosmological parameters $H_0=70\ km\ s^{-1}\ Mpc^{-1}$, $\Omega_m = 0.3$, and $\Omega_\Lambda = 0.7$. Uncertainties are quoted at the 68\% confidence level. The abundances are given with respect to the proto-Solar abundances of \citet{Lodders_2009}.

\section{Observations and Data Reduction}
\label{sec:Observations and Data Reduction}

We make use of \textit{XRISM}/Resolve observations, complemented by archival Chandra and XMM-Newton/EPIC data, to obtain a detailed view of the chemical composition of the Virgo cluster. The observations employed in this study are outlined below, and the following section describes the corresponding data analysis.

\begin{figure*}
    \centering
    \includegraphics[width=1\linewidth]{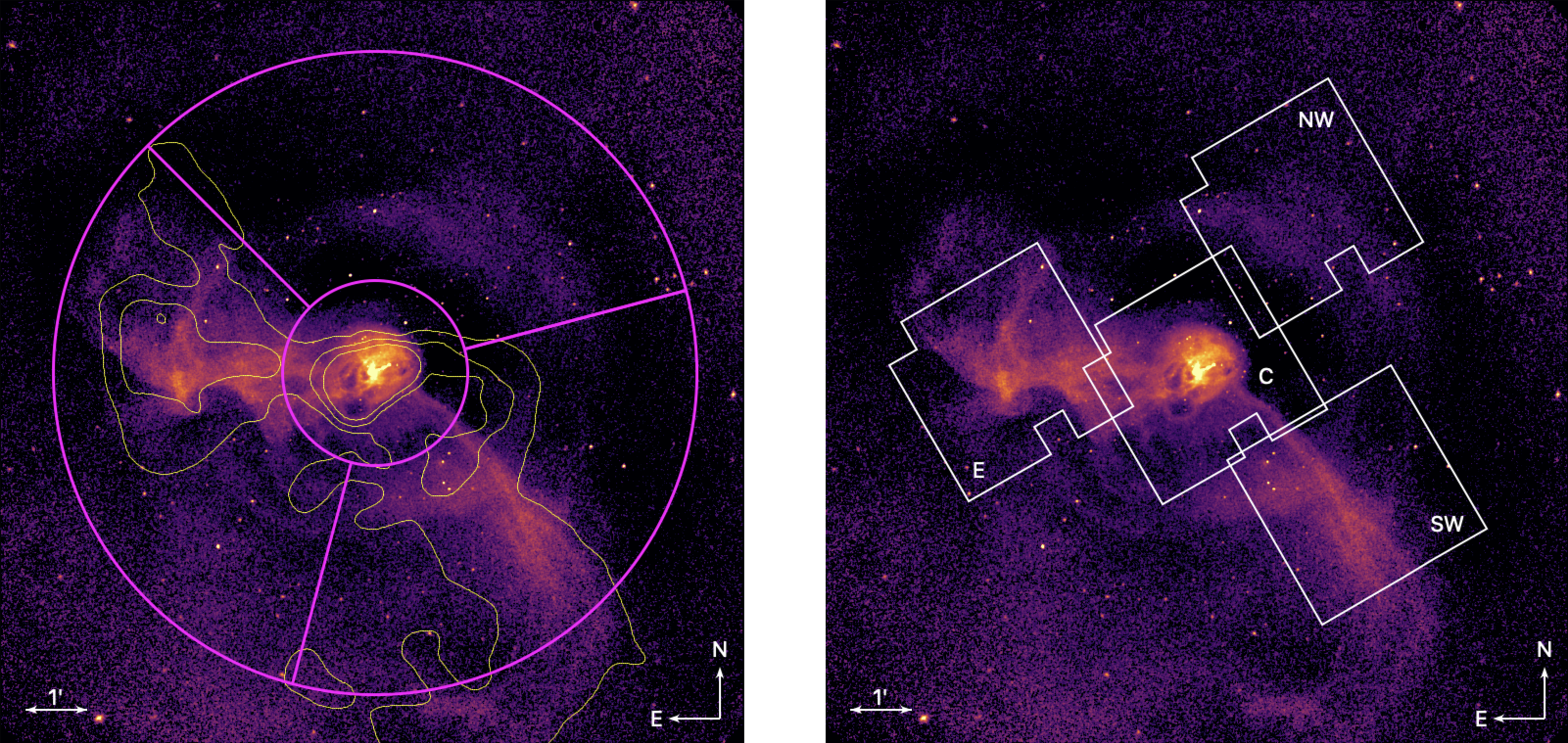}
    \caption{Chandra images of M87 in the 0.5-7 keV band, divided by a spherically symmetric $\beta$-model, smoothed with a Gaussian kernel of radius and sigma of 2 pixels each, and square-root scaled to enhance low-surface-brightness features. \textit{Left}: The four sky regions used for ray-tracing and \texttt{arf} generation for the \textit{XRISM}/Resolve FOVs are illustrated by the magenta shapes, with the corresponding detector regions shown in white on the right plot. The yellow contours highlight the radio emission at $\sim$4.6–4.9 GHz from a VLA interferometric image of M87 observed in 1986. \textit{Right}: The four \textit{XRISM}/Resolve FOVs.}
    \label{fig:M87_residual}
\end{figure*}

\subsection{\textit{XRISM} Observations}

Four distinct regions of M87 were observed in 2024 using the Resolve instrument onboard \textit{XRISM}, with the filter wheel in the open configuration and the gate valve closed in front of it. The central (hereafter C, see Fig~\ref{fig:M87_residual}) region of the cluster was observed on June 13 for a cleaned effective time of 116 ks (RA = 187.704, DEC = 12.390, ObsID: 300014010); the northwestern (NW) region was observed on May 26 and 27, for 26 ks and 143 ks, respectively (RA = 187.678, DEC = 12.435, ObsIDs: 300016010 and 300016020); the eastern (E) region was observed on May 30, for 160 ks (RA = 187.759, DEC = 12.391, ObsID: 300015010); the southwestern (SW) region was observed on June 2 and December 9, for 76 ks and 82 ks, respectively (RA = 187.661, DEC = 12.357, and RA = 187.660, DEC = 12.358, ObsIDs: 300017010 and 300017020), with the second SW observation (SW2) being rotated by 180° with respect to the first one (SW1). 

These pointings allow the probing of spatial variations in the chemical abundances within the central ICM of the Virgo cluster. The C field-of-view (FOV) targets the AGN, jet, and the directly surrounding ICM of M87, providing insight into the central metal enrichment and the impact of AGN feedback on the surrounding ICM. The E and SW regions cover the X-ray bright arms, which are the most striking features driven by the AGN activity \citep{Feigelson_1987}. They are believed to be uplifted multi-temperature \citep{Werner_2010} structures associated with AGN feedback \citep{Simionescu_2008,Million_2010}. Measuring abundances in these regions thus helps to trace the transport of metals away from the cluster core. In contrast, the NW pointing samples a region outside the bright X-ray arms that intersects the AGN-driven shock front \citep{Forman_2005, Million_2010}, providing a reference baseline for comparison with the strongly uplifted arm regions. Together, these observations allow a spatially resolved study of the enrichment history and metal transport mechanisms across different dynamical environments within the cluster.
\subsection{XRISM Data Reduction} 
\label{sec:XRISM Data Reduction}

The \textit{XRISM} data were processed following the methodology outlined in \citet{XRISM_2025} (hereafter \citetalias{XRISM_2025}), using the public version of \texttt{heasoft} 6.34 and CALDB version 10. Event screening was performed in accordance with the \textit{XRISM} ABC Guide v1.0 \footnote{\url{https://heasarc.gsfc.nasa.gov/docs/xrism/analysis/abc_guide/xrism_abc.html}}.
In short, we applied standard event filters based on the RISE\_TIME, DERIV\_MAX, and STATUS flags to reject non-X-ray events and false detection. Pixels 11 and 27 exhibited energy-scale jumps during ObsID 300016020 and in all observations, respectively. As this effect cannot be corrected \citep{Porter_2025}, they were excluded from the affected observations. 

Spectra were extracted over the full FOV of each pointing and restricted to high-resolution primary (Hp) events. Ancillary response files (\texttt{ARFs}) were computed in two configurations, IMAGE and POINT-SOURCE, to model the different emission components present in the \textit{XRISM}/Resolve fields. The Image \texttt{ARFs}, representing the diffuse emission from the ICM, were generated using the \texttt{xaarfgen} tool in IMAGE mode, with an exposure-corrected Chandra ACIS image of M87 in the 3.0–7.0 keV band as input. The AGN and the small-scale jet, visible on the Chandra image in the core of the C pointing, were masked on the image and filled with an average surface brightness derived from the surrounding pixels (see \citetalias{XRISM_2025} for details). Point-source \texttt{ARFs} were computed using \texttt{xaarfgen} in POINT SOURCE mode to account for the AGN, its jet, and low-mass X-ray binaries (LMXBs). The corresponding redistribution matrix files (\texttt{RMFs}) and exposure maps were generated with the \texttt{rslmkrmf} and \texttt{xaexpmap} tools, respectively. We created the \texttt{RMFs} from cleaned event files that included all valid event grades except low-resolution secondary (Ls), due to the large number of false Ls events associated with the on-board application of secondary-pulse detection to atypically shaped pulses such as clipped events. For the count rates of these observations, the high-primary (Hp) branching ratio is more accurately determined after the elimination of all Ls events. These response files were used consistently across all spectral fits to ensure accurate modeling of both the spatially extended ICM and the localized sources.
The spectra of the non X-ray background (NXB), i.e. the background noise that is not caused by celestial X-ray sources, were extracted from the Night-Earth observation database using the \texttt{rslnxbgen} task. We selected a time interval from \texttt{timefirst=}-730 to \texttt{timelast}=150, with no cutting on CORTIME. As the SW region (ObsID 17020) was observed for a second time 6 months after all the other M87 observations, its data were analyzed with an NXB spectrum extracted from an extended Night-Earth database. We produced one NXB spectrum per M87 observation, selected with the same filtering criteria as applied to the M87 data. 

\subsection{Archival Chandra \& EPIC Observations}
\label{sec:Archival Chandra EPIC Observations}

We utilized the full field Chandra observations of M87 taken on ACIS-S and ACIS-I in 2000, 2002 and 2005 \citep[Obs. IDs 352, 2707, 3717, 5826, 5827, 5828, 6186, 7210, 7211 and 7212;][]{DiMatteo_2003,Forman_2005}. For details of the data reprocessing, see \citet{russell2018} and \citetalias{XRISM_2025}. Time periods affected by flares were excluded and the resulting total exposure time was $608\,$ks. Extended emission from M87 covers the entire Chandra detector so appropriate blank sky backgrounds were generated for each observation from the blank sky datasets and appropriate responses were generated. Spectra were extracted from each observation using regions corresponding to the different \textit{XRISM} pointings and were grouped with a minimum of 1 count per bin.

We also make use of XMM-Newton/EPIC observations (Obs. IDs 0803670501 and 0803670601). These were re-analyzed with SAS v22 and using the pipeline described in \citet{Rossetti_2024}. The net exposure time is $185\,$ks for the MOS1/2 and $140\,$ks for the pn detectors. Similarly to the Chandra data, spectra were extracted for both observations from regions closely matched to the \textit{XRISM} fields of view. 

\section{Data Analysis}
\label{sec:Data analysis}

\subsection{Spectral modeling}
\label{sec:Spectral modeling}

The analysis focuses on full-FOV spectra of regions C, E, NW and SW of the Virgo cluster centered on M87 (see Fig. \ref{fig:M87_residual}, right). They were extracted and analyzed in the 1.7-11 keV range where Resolve achieves a spectral resolution of $\sim$5 eV. Spectral fitting was performed with \texttt{SPEX} \citep{Kaastra_2024} in combination with the \texttt{SPEXACT} atomic database (v3.08.02). The spectra were optimally binned using the \texttt{obin} algorithm \citep{Kaastra_Bleeker_2016}, and the best-fit parameters were obtained by minimizing the C-statistic \citep{cash1979}. This work also includes specific results from \citet{Simionescu_2026} (hereafter \citetalias{Simionescu_2026}) obtained with \texttt{Xspec} \citep[v12.15.0][]{Xspec}. 

To investigate the thermal structure of the ICM in each region, we explored a set of spectral models of increasing complexity. Temperature inhomogeneities in the ICM are expected due to multiphase gas, projection along the line of sight, and mixing driven by AGN feedback and large-scale motions. In addition, the limited low-energy sensitivity of Resolve may hinder the detection of cooler plasma components. To assess the robustness of the inferred abundances and quantify potential biases from unresolved thermal structure, we tested single-temperature, multi-temperature, and differential emission measure (EM) models. All models consist of redshifted (\texttt{REDS}) and absorbed (\texttt{HOT}) optically thin thermal plasma emission in collisional ionization equilibrium with velocity broadening (\texttt{CIE}). The redshift was left free in all models. The Galactic hydrogen density of the neutral interstellar matter was fixed at 1.26$\times 10^{20}$ cm$^{-2}$, corresponding to the weighted average of neutral and molecular hydrogen (\citet{HI4PI_2016}). Thermal equilibrium was assumed by coupling the electron temperature \texttt{t} to the ion temperature \texttt{it} in CIE components. The source distance was set to 16.4 Mpc \citep{Rackers_2024}. We describe below the tested models and fitting methodology, with parameters in \texttt{monospace font}. 

\paragraph{1T -- One-temperature model:} \texttt{REDS $\times$ HOT $\times$ CIE}. A single-temperature plasma was fitted in all four regions. Free parameters include electron temperature \texttt{t}, emission measure \texttt{norm}, turbulent velocity \texttt{vrms}, and abundances of Si, S, Ar, Ca, Cr, Fe and Ni (\texttt{14, 16, 18, 20, 24, 26} and \texttt{28}). Abundances of H, He, Li, Be, B were fixed to 1. The reference atom parameter was set to hydrogen, such that all abundances are defined relative to H in Solar units. All other elements were tied to Fe. 
\paragraph{2T - Two-temperature model:} \texttt{HOT $\times$ (REDS$_{hot}$ $\times$ CIE$_{hot}$ + REDS$_{cold}$ $\times$ CIE$_{cold}$)}. A second thermal component is introduced to capture unresolved multiphase temperature structure, which has already been observed in the X-ray arms \citep[e.g.][]{Belsole_2001,Molendi_2002,Matsushita_2002, Simionescu_2008}. This model was unconstrained in the NW region so results are shown only for C, E and SW. Free parameters are the temperatures, the \texttt{norm} of each gas phases, and a common turbulent velocity \texttt{vrms}. Abundances of Si, S, Ar, Ca, Cr, Fe and Ni were tied between the CIEs and left free; remaining elements were treated as in the 1T model. For SW, the hot (cold) emission measures were tied between SW1 and SW2.
\paragraph{2T-2Z - Two-temperature model with independent abundances:} Applied to E and SW, this model follows the 2T setup, with different abundances of Si, S, Ar, Ca, Cr, Fe and Ni between the hot and cold phases. However, the data cannot independently constrain the abundances of both phases, and since the cold phase in the X-ray bright arms originates from M87, we assume it shares the same composition. Accordingly, abundances of Si, S, Ar, Ca, Cr, Fe, and Ni are left free in the hot phase, while those of the cold phase are fixed to the 2T model best-fit values from region C. The results and implications of this model are discussed in \ref{sec:Chemical properties hot vs. cool gas}.
\paragraph{2T Chandra - Two-temperature model with Chandra constraints:} The cold-phase temperature and cold-to-hot EM ratio were fixed to values derived from a 2T fit to Chandra data (Sec. \ref{sec:Archival Chandra EPIC Observations}). Theses spectra, fitted in the 0.5-7 keV band with \texttt{TBabs*(vapec+vapec)} in \texttt{Xspec} (v12.15.0, AtomDB 3.1.3), had temperatures, redshifts, and Fe abundance free, while other parameters were tied between the components.
\paragraph{2T EPIC - Two-temperature model with \texttt{EPIC} constraints:} This model uses the same setup as 2T Chandra, with constraints from XMM-Newton/EPIC spectra fitted with \texttt{TBabs*(vapec+vapec)+pow}, including sky and detector background components. Free abundances of O, Ne, Mg, Si, S, Ar, Ca, Fe and Ni were tied between thermal components. Consistent with previous EPIC studies \citep{Belsole_2001,Matsushita_2002,Molendi_2002}, these fits suggest cooler gas phases than those inferred from \texttt{XRISM}, due to the Fe-L line sensitivity in the \texttt{EPIC} soft band, which is inaccessible to Resolve with the gate valve closed. These last two models were tested to account for this cooler phase. 
\paragraph{Wdem - Differential emission measure (DEM) model:} \texttt{REDS $\times$ HOT $\times$ WDEM}. This multi-temperature model describes continuous thermal structure in the cluster core, assuming a power-law differential emission measure between $\beta T_{max}$ and $T_{max}$. The slope controls the weighting of emission across temperatures: large positive slopes approach a single-temperature distribution, while smaller or negative slopes indicate a broader contribution from cooler gas. The slope and $\beta$ parameters are left free to vary; all other parameters settings are identical to those adopted in the 1T model. 
\paragraph{Gaussian Lin(T) and Log(T) - DEM models:} These models generalize the 1T setup to include a Gaussian temperature distribution. The EM distribution follows $Y(x) = \frac{Y_0}{\sigma_T \sqrt{2\pi}} \exp\left( -\frac{(x - x_{\text{mean}})^2}{2\sigma_T^2} \right)$, with $x = kT$ in Lin(T), and $x = log(kT)$ in Log(T). $kT_{mean}$ is the mean temperature of the distribution, $\sigma_T$ the width of the distribution, and $Y_0$ the total integrated emission measure. $kT_{mean}$ and \texttt{$\sigma_T$} are free to vary; all other parameters settings are identical to those adopted in the 1T model. 
\paragraph{CLUS:} \texttt{REDS $\times$ HOT $\times$ CLUS}. This model was applied to the NW region only. The CLUS component was initialized using the best-fit 3D radial gas density profiles published in \citetalias{XRISM_2025}, the temperature and Fe abundance profiles derived from deprojected Chandra and Suzaku data \citep{Simionescu_2017_Suzaku}, following the formalism introduced in \citet{Stefanova_2025}. The profiles, their fits and best-fit parameters are reported in Appendix~\ref{CLUS parameters}. The free parameters are the redshift \texttt{z}, the temperature \texttt{$T_h$}, the density $n_1$ and the turbulent velocity $a_v$. Elemental abundance are treated the same way as in the 1T model. 
\paragraph{Xspec:} This model refers to \texttt{Model B} in \citetalias{Simionescu_2026}. The full-FOV spectra were fitted over the 1.7–9.0 keV band with a two-temperature plasma model using \texttt{Xspec} and AtomDB v3.1.2. In region C, the model follows the 2T-model setup. In the offset pointings (E, SW, NW), the spectra are fitted simultaneously, with some properties of the cooler gas (temperature and velocity broadening) coupled between the E and SW pointings, and the NW region assumed to be single phase. The abundances of Si, S, Ar, Ca, Cr, Fe, and Ni are fitted independently in each offset pointing; where two thermal components are present, the abundances of the two phases are coupled to each other.\\

The NW pointings were fitted simultaneously, with all ICM parameters tied between the datasets. For all except the 2T models, all ICM parameters were tied between the SW datasets, fitted simultaneously, except for the emission measures. The redshift of SW2 was blueshifted by $\sim$55 $\mathrm{km.s}^{-1}$ relative to SW1 to account for the difference in heliocentric velocity (see \citetalias{Simionescu_2026} for details). 
The AGN and jet emission is modeled as an absorbed power-law with the photon index fixed at $\Gamma=2.35$ and the 2-7 keV unabsorbed flux fixed at $1.52 \times 10^{-12}\ \textrm{erg}/\textrm{cm}^2/\textrm{s}$, based on Chandra observations in which the AGN and jet are spatially resolved. Given the stability of the AGN+jet spectrum \citep{russell2018}, these parameters were kept fixed in all fits (see \citetalias{XRISM_2025} for details). LMXB emission within each region is accounted for with a separate absorbed power-law with photon index fixed to $1.65$ \citep[e.g.][]{irwin2003,Revnivtsev2014} and free 2-7 keV flux. 
We computed one NXB spectrum per source observation (see Sec. \ref{sec:XRISM Data Reduction}). Each NXB spectrum was fitted individually in the 1.7-11 keV energy range using an empirical NXB model, designed to describe the instrumental background. The model consists of a power-law continuum, 17 Gaussian components approximating the instrumental emission lines, and 3 scaling factors. The fitting was performed with the supplied diagonal \texttt{rmf} and proceeded in two steps: first, the overall normalization of the model was fitted and then kept frozen; secondly, the normalization of all Gaussian components were adjusted simultaneously. The best-fit NXB models were simulated to create mock NXB spectra using the \texttt{HEASOFT} \texttt{fakeit} command, excluding Poisson noise, assuming an arbitrarily large exposure to maximize the statistics, and folded through the diagonal \texttt{rmf}. For each M87 observation, the corresponding simulated NXB spectrum was used as the background file in the \texttt{SPEX} spectral fits. 

\subsection{Spatial-Spectral Mixing}
\textit{XRISM}/Resolve has a FOV of approximately $3'\times3'$ and its Point Spread Function (PSF) has a Half Power Diameter (HPD) of $1.3'$ \citep{Takayuki_2024}, where the HPD denotes the diameter of the circular region enclosing 50\% of the photons from a point source. Although the HPD is smaller than the FOV, it is comparable in scale, implying that a significant fraction of photons originating outside a given extraction region leaks into it through the extended wings of the PSF. The X-ray spectra extracted from individual detector regions are thus subject to contamination from photons originating from nearby regions.

This effect, referred to as Spatial-Spectral Mixing (SSM)\footnote{https://heasarc.gsfc.nasa.gov/docs/xrism/proposals/POG/}, produces cross-contamination between neighboring sky regions. The spectrum extracted from any detector region is therefore a spatially weighted superposition of emission from multiple sky regions. Accounting for SSM is essential for accurate measurements of spatially varying spectral parameters, including abundances, and has been demonstrated to be important for Resolve analyses of extended sources \citep{Hitomi_2018, XRISM_2025}. To model it, we perform ray-tracing simulations with the \texttt{xrtraytrace} task, which propagates photons from an assumed sky brightness distribution through the telescope optics, to the detector plane, and calculates the transfer of photons between sky and detector regions. The input sky distribution is the same exposure-corrected 3.0-7.0 keV Chandra ACIS image of M87 used to generate the extended \texttt{ARFs} (Sect. \ref{sec:XRISM Data Reduction}) and used in both \citetalias{XRISM_2025} and \citetalias{Simionescu_2026}. This image provides a sub-arcsecond-resolution map of the X-ray surface brightness distribution of M87, and serves as a high-fidelity template for the photon spatial distribution. The four sky regions used for ray-tracing are show in magenta on Fig.~\ref{fig:M87_residual}.

The ray-tracing estimates indicate that the dominant contamination of the outer pointings (E, NW, SW) arises from the central region that contributed more than 5\% of their total detected flux. Consequently, in the spectral modeling of regions E, NW and SW, we include contaminating emission originating only from the region C. This emission is modeled using the best-fit 2T model applied to region C's full FOV spectrum, with additional power-laws accounting for the AGN, inner jet, and LMXBs, and all parameters are fixed throughout the analysis. 

\section{Results}
\label{sec:Results}

Relative abundances are generally less sensitive to systematic uncertainties than absolute metal abundances, making ratios with respect to Fe (X/Fe) more reliable diagnostics. We first examine how different temperature-structure models affect these measurements, then assess the impact of using different atomic databases. Finally, we present and discuss the values obtained with a chosen fiducial model. A detailed analysis of instrumental systematics on the abundance ratios is provided in Appendix~\ref{sec:Appendix_A_systematic_tests}. 

\subsection{Impact of temperature structure}
\label{sec:Impact of temperature structure}

The best-fit abundances obtained from the 1.7-11 keV spectral analysis using the models presented in Sect. \ref{sec:Spectral modeling} are reported in tables \ref{tab:best_fit_and_abundance_ratios_NW}, \ref{tab:best_fit_and_abundance_ratios_C}, \ref{tab:best_fit_and_abundance_ratios_E}, and \ref{tab:best_fit_and_abundance_ratios_SW}. Other best-fit parameters are reported in appendix \ref{Best fit parameters}. The inferred temperatures are comparable with previous X-ray measurements of M87 at similar projected distances obtained with different instruments, including XMM-Newton \citep{Belsole_2001,Molendi_2002,Matsushita_2002,Simionescu_2008}, Chandra \citep{Young_2002}, and Suzaku \citep{Simionescu_2017_Suzaku}. A dedicated and detailed analysis of the temperature structure in the four regions using \textit{XRISM}/Resolve data will be presented in Martin, J. et al (in prep). The line-of-sight velocities measured over the full-field regions are consistent with values obtained in sub-regions of the C and NW pointings in \citetalias{XRISM_2025}, as well as in regions C, E and SW in \citetalias{Simionescu_2026}, where these measurements are discussed in greater detail. Similarly, constrained values of redshifts and LMXB fluxes in the 2-7 keV band are consistent with those reported in \citetalias{Simionescu_2026}. We note that in the SW, we were unable to obtain a well-constrained LMXB flux, and a 2$\sigma$ upper limit is derived for each model instead. 

\begin{figure*}
    \centering
    \includegraphics[width=1\linewidth]{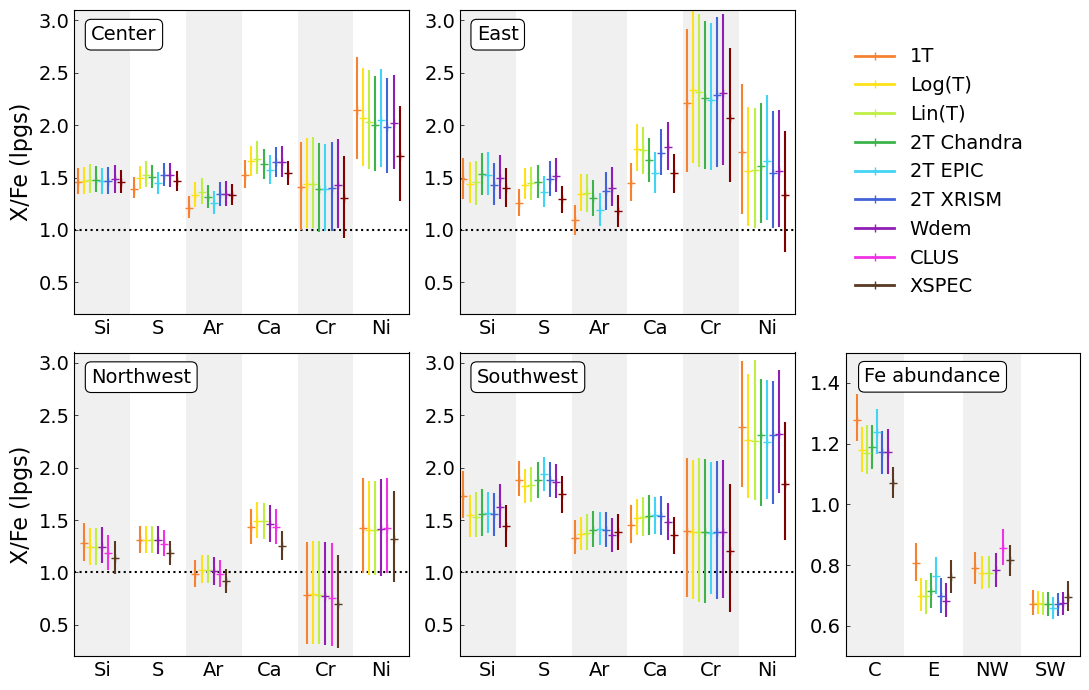}
    \caption{Abundance ratios of Si, S, Ar, Ca, Cr, and Ni with respect to Fe inferred from the \textit{XRISM}/Resolve data in the four regions (indicated in each panel). The horizontal dashed line indicates a ratio of 1. The bottom right sub-figure shows the absolute Fe abundance per region. The values are derived from the different spectral models described in Sect.\ref{sec:Spectral modeling} and indicated with different colors, fitted to the full field-of-view, 1.7-11 keV spectra using \texttt{SPEX} version 3.08.02. and \texttt{Xspec} version 12.15.0.}
    \label{fig:Abundances_and_abundances_ratios}
\end{figure*}

Fig. \ref{fig:Abundances_and_abundances_ratios} presents the best-fit absolute Fe abundance together with the abundance ratios of the $\alpha$ (Si, S, Ca, Ar) and Fe-peak (Ni, Cr) elements relative to Fe, as constrained with the different modeling strategies. The corresponding values are listed in Tables \ref{tab:best_fit_and_abundance_ratios_NW}, \ref{tab:best_fit_and_abundance_ratios_C}, \ref{tab:best_fit_and_abundance_ratios_E}, and \ref{tab:best_fit_and_abundance_ratios_SW}. While the absolute Fe abundance shows some sensitivity to the assumed temperature structure, the inferred abundance ratios are remarkably stable. In each of the four regions, the X/Fe ratios derived from 1T, 2T, and DEM models are mutually consistent within their statistical uncertainties, indicating that the abundance ratios are robust against the specific choice of temperature parameterization. In regions C and E, the 1T model yields slightly higher Fe abundances than the multi-temperature descriptions, although the difference remains consistent within $1\sigma$. This leads to correspondingly lower abundance ratios of the $\alpha$-elements in these regions. Given the 1.7–11 keV energy band of our analysis, the Fe-L complex ($\sim$0.7–1.2 keV) lies outside the fitted band-pass, such that the abundance constraints are driven primarily by Fe-K and $\alpha$-element emission lines. Although temperature–abundance degeneracies are weaker in this energy range than in the Fe–L band, they are not negligible in the presence of multi-phase gas. In particular, broader or hotter temperature distributions can alter line emissivity ratios and bias single-temperature fits toward higher Fe abundances, which we interpret as the origin of this observed offset in regions C and E. 

Additionally, the goodness-of-fit statistics, also reported in in Tables~\ref{tab:best_fit_and_abundance_ratios_NW}, \ref{tab:best_fit_and_abundance_ratios_C}, \ref{tab:best_fit_and_abundance_ratios_E}, and \ref{tab:best_fit_and_abundance_ratios_SW}, show that the \textit{XRISM}/Resolve spectra strongly disfavor a single-temperature description across all pointings, and instead require the presence of temperature structures, either intrinsic or induced by projection effects. In regions C, E and SW, where the ICM is known to be multi-phase (e.g. \citet{Sakelliou_2002}, \citet{Simionescu_2008}, \citet{Werner_2013}), the 1T model biases Fe estimates, by adjusting the Fe abundance to compensate for unresolved temperature structure. In these three regions, both differential emission measure and discrete multi-temperature models provide statistically significant and comparable improvements over the 1T model, demonstrating that temperature broadening is robustly detected, although its detailed parameterization is weakly constrained. By contrast, in the NW region, the improvement of differential emission measure over 1T can be naturally attributed to projection through smooth radial temperature gradients rather than to intrinsically multiphase gas. 
The consistency of the inferred X/Fe abundance ratios across all tested thermal models demonstrates that the chemical enrichment pattern in each region is robustly constrained by the \textit{XRISM}/Resolve data, despite uncertainties in the detailed temperature structure. 

\subsection{Impact of atomic database}
\label{sec:Impact of atomic database}

Different atomic databases predict in some cases different emissivities for individual emission lines. Therefore, to assess the dependence of the abundance measurements on the choice of atomic database --\texttt{SPEXACT} with \texttt{SPEX} vs. \texttt{ATOMDB} with \texttt{Xspec}-- we compare our results, obtained with \texttt{SPEXACT}, to the values derived with Model B from \citetalias{Simionescu_2026}, that we refer to as the "1T" or "2T" \texttt{Xspec} models, for the NW and C, E and SW, respectively.

The abundances derived with the 2T model, shown in dark blue for \texttt{SPEX} and brown for \texttt{Xspec}, and the 1T model in NW, in orange for \texttt{SPEX} and brown for \texttt{Xspec} on Fig. \ref{fig:Abundances_and_abundances_ratios}, are consistent within uncertainties for most elements and regions. A notable exception is the Fe abundance in region C, with the 2T \texttt{Xspec} analysis being systematically lower by 9.6\%\footnote{This value is obtained by averaging the fractional differences between the \texttt{Xspec} Fe abundance and each of the corresponding multi-temperature \texttt{SPEX} model measurements in region C.} relative to the \texttt{SPEX} multi-temperature models, leading to a statistically significant discrepancy. This behavior does however not impact the ratios and is not observed in the other pointings. This difference likely arises from a combination of two effects. First, region C contains a larger fraction of relatively cool plasma compared to the other pointings. In this temperature regime, Fe XXII and Fe XXIII line complexes around $\sim$6.6–6.7 keV contribute significantly to the Fe-K emission. Accurate modeling of these ions is therefore particularly important since differences in the treatment of Fe-L and Fe-K satellite lines between \texttt{SPEXACT} and \texttt{ATOMDB} can lead to systematic shifts in the inferred Fe abundance \citep{De_Plaa_2017, Hitomi_2017}. Second, the best-fit LMXB flux in this region differs between the two analyses at about 2$\sigma$ level. Indeed, the 2T \texttt{Xspec} fit yields a logarithmic LMXB flux in 2-7 keV in units of erg.cm$^{-2}$.s$^{-1}$ of $\log( F_{LMXB}) = -11.81 \pm 0.05$, while the 2T \texttt{SPEX} fit yields $\log(F_{LMXB}) = -11.92 \pm 0.05$. This difference is likely driven by the different energy bands used in the fits (1.7–9 keV with \texttt{Xspec} vs. 1.7–11 keV with \texttt{SPEX}), which affect the separation between thermal and non-thermal continuum components. The resulting continuum differences propagate directly into the Fe equivalent width and hence the inferred abundance. 

This comparison shows that atomic data systematics introduce only modest variations in the inferred abundance ratios. Other systematic uncertainties including LMXB fluxes, instrument calibration, and PSF contamination, are also small compared to the statistical errors, as shown in detail in Appendix~\ref{sec:Appendix_A_systematic_tests}. 

\subsection{Final abundance ratios}

Since the data require a spread in temperature and the precise form of the temperature distribution remains unconstrained (Sect. \ref{sec:Impact of temperature structure}), we need to adopt a well-defined fiducial model for the determination of the final abundance ratios. We therefore chose the Gaussian logarithmic temperature distribution (Log(T)) model as the fiducial description. It provides a statistically significant improvement over a single-temperature model while introducing only a single additional free parameter. Its differential emission measure is physically interpretable as a spread of plasma temperatures, and it is fully constrained by the \textit{XRISM} spectra without relying on external soft-band priors. The alternative tested multi-temperature descriptions yield comparable fit quality and support the same physical conclusion. This choice is further supported by the work of \citet{Chatzigiannakis_2026} where the analysis of mock \textit{XRISM} observations of TNG-Cluster concluded that a lognormal temperature distribution better described the thermal structure of the ICM. 

\begin{figure}
    \centering
    \includegraphics[width=1\linewidth]{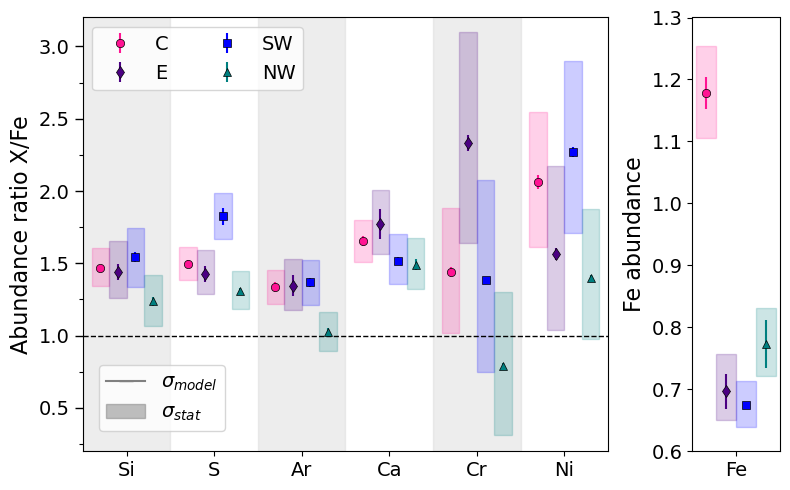}
    \caption{Abundance ratios (\textit{left}) and absolute Fe abundances (\textit{right}) measured with \textit{XRISM}/Resolve in C, E, SW, and NW using the fiducial gdem–Log(T) model. Colored rectangles indicate statistical uncertainties relative to the fiducial model while vertical error bars show model systematic uncertainties from the dispersion of alternative models (excluding 1T in C, E and SW).}
    \label{fig:Abundance_ratios_per_pointing}
\end{figure}

Figure \ref{fig:Abundance_ratios_per_pointing} compares the abundance ratios derived with the reference Log(T) model across the four regions. The dispersion of the best-fit abundances with respect to this model, $\sigma_{\mathrm{model}}$, obtained with the alternative thermal descriptions is treated as a systematic uncertainty, as it reflects model dependence rather than statistical noise. As the 1T parametrization is a poor representation of the ICM in regions C, E and SW, it was excluded from the calculations. The colored rectangles indicate the $1\sigma$ statistical uncertainties associated with the fiducial model, while the vertical error bars represent the uncertainties due to the different modeling strategies with respect to the Log(T) model, which were computed as follows. With $X_{ref}$ the abundance ratio obtained with the Log(T) model, and $N$ the number of alternative models yielding values $X_i$, the model-induced dispersion is defined as:
$\sigma_{\mathrm{model}} = \left[\frac{1}{N} \sum_{i=1}^{N} \left( X_i - X_{\mathrm{ref}} \right)^2 \right]^{1/2}$. This figure highlights one of the key results of this work: the abundance ratios of all measured elements are found to be super-Solar in the four regions of M87 mapped by \textit{XRISM}/Resolve. Ar/Fe in the NW region, and Cr/Fe in NW and SW, are the only ratios consistent with Solar values. 

We also note that all elements are detected at a significance level above 3$\sigma$ with the following exception: Ni in the E and Cr in the SW, which are detected at the 3$\sigma$ and 2$\sigma$ levels respectively, and Cr in the NW, which is detected at less than 2$\sigma$. The differing significance levels of the Cr and Ni abundances in the offset pointings are likely due to statistical noise. 

Overall, for all elements and pointings, uncertainties associated with the modeling strategy are marginal compared to the statistical errors. Variations in the assumed ICM temperature structure therefore produce negligible changes in the best-fit abundance ratios. This behavior indicates that the abundance ratios are not strongly degenerate with the details of the ICM thermal modeling, in agreement with the conclusions of Sect.~\ref{sec:Impact of atomic database} and also demonstrates that the Log(T) model provides a robust framework for measuring abundance ratios.

\begin{table*}[ht]
\renewcommand{\arraystretch}{1.3}
\centering
\caption{Best-fit parameters from the spectral fitting of the NW region data in the 1.7-11 keV band.}
\begin{tabular}{lcccccc}
\hline
Parameter & 1T & 1T Xspec & Log(T) & Lin(T) & Wdem & CLUS \\
\hline
Si & $1.01^{+0.14}_{-0.12}$ & $0.93^{+0.12}_{-0.11}$ & $0.96^{+0.12}_{-0.11}$ & $0.96^{+0.13}_{-0.11}$ & $0.97^{+0.13}_{-0.10}$ & $1.02^{+0.13}_{-0.12}$ \\
S & $1.03^{+0.08}_{-0.07}$ & $0.97^{+0.07}_{-0.07}$ & $1.01^{+0.08}_{-0.07}$ & $1.02^{+0.08}_{-0.07}$ & $1.03^{+0.08}_{-0.07}$ & $1.09^{+0.08}_{-0.08}$ \\
Ar & $0.78^{+0.09}_{-0.08}$ & $0.75^{+0.08}_{-0.08}$ & $0.79^{+0.09}_{-0.08}$ & $0.79^{+0.09}_{-0.08}$ & $0.79^{+0.09}_{-0.09}$ & $0.84^{+0.09}_{-0.09}$ \\
Ca & $1.13^{+0.11}_{-0.10}$ & $1.02^{+0.10}_{-0.09}$ & $1.15^{+0.11}_{-0.10}$ & $1.15^{+0.11}_{-0.10}$ & $1.15^{+0.11}_{-0.10}$ & $1.23^{+0.12}_{-0.11}$ \\
Cr & $0.62^{+0.40}_{-0.37}$ & $0.57^{+0.38}_{-0.34}$ & $0.61^{+0.39}_{-0.37}$ & $0.61^{+0.40}_{-0.36}$ & $0.61^{+0.40}_{-0.36}$ & $0.64^{+0.45}_{-0.39}$ \\
Fe & $0.79^{+0.05}_{-0.05}$ & $0.82^{+0.05}_{-0.05}$ & $0.77^{+0.06}_{-0.05}$ & $0.77^{+0.06}_{-0.05}$ & $0.78^{+0.06}_{-0.06}$ & $0.86^{+0.06}_{-0.06}$ \\
Ni & $1.13^{+0.37}_{-0.33}$ & $1.08^{+0.37}_{-0.33}$ & $1.08^{+0.36}_{-0.32}$ & $1.09^{+0.36}_{-0.33}$ & $1.10^{+0.37}_{-0.34}$ & $1.22^{+0.40}_{-0.36}$ \\
\hline
\end{tabular}
\label{tab:best_fit_and_abundance_ratios_NW}
\end{table*}

\begin{table*}[ht]
\renewcommand{\arraystretch}{1.3}
\centering
\caption{Best-fit parameters from the spectral fitting of the C region data in the 1.7-11 keV band.}
\begin{tabular}{lcccccccc}
\hline
Parameter & 1T & Log(T) & Lin(T) & Wdem & 2T Chandra & 2T EPIC & 2T & 2T Xspec \\
\hline
Si & $1.86^{+0.11}_{-0.11}$ & $1.73^{+0.11}_{-0.10}$ & $1.73^{+0.11}_{-0.10}$ & $1.74^{+0.11}_{-0.10}$ & $1.75^{+0.10}_{-0.10}$ & $1.82^{+0.11}_{-0.10}$ & $1.72^{+0.11}_{-0.10}$ & $1.56^{+0.09}_{-0.09}$ \\
S & $1.78^{+0.07}_{-0.07}$ & $1.76^{+0.07}_{-0.07}$ & $1.78^{+0.07}_{-0.07}$ & $1.78^{+0.07}_{-0.07}$ & $1.79^{+0.07}_{-0.07}$ & $1.79^{+0.07}_{-0.07}$ & $1.79^{+0.07}_{-0.07}$ & $1.57^{+0.06}_{-0.06}$ \\
Ar & $1.55^{+0.10}_{-0.09}$ & $1.57^{+0.10}_{-0.09}$ & $1.59^{+0.10}_{-0.09}$ & $1.57^{+0.09}_{-0.09}$ & $1.56^{+0.10}_{-0.09}$ & $1.56^{+0.10}_{-0.09}$ & $1.57^{+0.10}_{-0.09}$ & $1.43^{+0.08}_{-0.08}$ \\
Ca & $1.95^{+0.12}_{-0.12}$ & $1.95^{+0.12}_{-0.12}$ & $1.95^{+0.12}_{-0.12}$ & $1.94^{+0.12}_{-0.12}$ & $1.93^{+0.12}_{-0.12}$ & $1.95^{+0.12}_{-0.12}$ & $1.93^{+0.12}_{-0.12}$ & $1.65^{+0.10}_{-0.09}$ \\
Cr & $1.80^{+0.54}_{-0.51}$ & $1.69^{+0.51}_{-0.48}$ & $1.68^{+0.51}_{-0.49}$ & $1.68^{+0.50}_{-0.48}$ & $1.65^{+0.51}_{-0.48}$ & $1.72^{+0.52}_{-0.49}$ & $1.64^{+0.51}_{-0.47}$ & $1.39^{+0.43}_{-0.40}$ \\
Fe & $1.28^{+0.08}_{-0.07}$ & $1.18^{+0.08}_{-0.07}$ & $1.17^{+0.09}_{-0.07}$ & $1.17^{+0.08}_{-0.07}$ & $1.19^{+0.07}_{-0.07}$ & $1.24^{+0.08}_{-0.07}$ & $1.17^{+0.07}_{-0.07}$ & $1.07^{+0.05}_{-0.05}$ \\
Ni & $2.74^{+0.62}_{-0.58}$ & $2.43^{+0.55}_{-0.51}$ & $2.37^{+0.54}_{-0.51}$ & $2.36^{+0.52}_{-0.49}$ & $2.37^{+0.54}_{-0.50}$ & $2.54^{+0.58}_{-0.54}$ & $2.32^{+0.53}_{-0.49}$ & $1.83^{+0.50}_{-0.46}$ \\
\hline
\end{tabular}
\label{tab:best_fit_and_abundance_ratios_C}
\end{table*}

\begin{table*}[ht]
\renewcommand{\arraystretch}{1.3}
\centering
\caption{Best-fit parameters from the spectral fitting of the E region data in the 1.7-11 keV band.}
\begin{tabular}{lcccccccc}
\hline
Parameter & 1T & Log(T) & Lin(T) & Wdem & 2T Chandra & 2T EPIC & 2T & 2T Xspec \\
\hline
Si & $1.20^{+0.13}_{-0.12}$ & $1.00^{+0.12}_{-0.11}$ & $1.02^{+0.12}_{-0.12}$ & $1.02^{+0.12}_{-0.11}$ & $1.09^{+0.12}_{-0.11}$ & $1.17^{+0.13}_{-0.12}$ & $1.00^{+0.12}_{-0.11}$ & $1.06^{+0.12}_{-0.11}$ \\
S & $1.01^{+0.07}_{-0.07}$ & $0.99^{+0.07}_{-0.07}$ & $1.01^{+0.08}_{-0.07}$ & $1.04^{+0.08}_{-0.07}$ & $1.04^{+0.08}_{-0.07}$ & $1.04^{+0.08}_{-0.07}$ & $1.03^{+0.08}_{-0.08}$ & $0.98^{+0.07}_{-0.07}$ \\
Ar & $0.88^{+0.10}_{-0.09}$ & $0.94^{+0.10}_{-0.10}$ & $0.94^{+0.11}_{-0.10}$ & $0.96^{+0.10}_{-0.10}$ & $0.93^{+0.10}_{-0.09}$ & $0.90^{+0.10}_{-0.09}$ & $0.95^{+0.10}_{-0.10}$ & $0.90^{+0.10}_{-0.09}$ \\
Ca & $1.17^{+0.12}_{-0.11}$ & $1.24^{+0.13}_{-0.12}$ & $1.23^{+0.13}_{-0.12}$ & $1.22^{+0.12}_{-0.12}$ & $1.19^{+0.12}_{-0.11}$ & $1.17^{+0.12}_{-0.11}$ & $1.21^{+0.12}_{-0.12}$ & $1.17^{+0.12}_{-0.11}$ \\
Cr & $1.78^{+0.56}_{-0.51}$ & $1.63^{+0.52}_{-0.47}$ & $1.61^{+0.51}_{-0.48}$ & $1.57^{+0.50}_{-0.46}$ & $1.61^{+0.51}_{-0.47}$ & $1.71^{+0.54}_{-0.49}$ & $1.60^{+0.50}_{-0.46}$ & $1.58^{+0.49}_{-0.45}$ \\
Fe & $0.81^{+0.07}_{-0.06}$ & $0.70^{+0.06}_{-0.05}$ & $0.70^{+0.05}_{-0.06}$ & $0.68^{+0.06}_{-0.05}$ & $0.72^{+0.06}_{-0.06}$ & $0.76^{+0.06}_{-0.06}$ & $0.70^{+0.06}_{-0.05}$ & $0.76^{+0.06}_{-0.05}$ \\
Ni & $1.41^{+0.52}_{-0.47}$ & $1.09^{+0.41}_{-0.36}$ & $1.10^{+0.40}_{-0.38}$ & $1.06^{+0.39}_{-0.35}$ & $1.15^{+0.43}_{-0.38}$ & $1.26^{+0.47}_{-0.42}$ & $1.08^{+0.40}_{-0.36}$ & $1.02^{+0.46}_{-0.41}$ \\
\hline
\end{tabular}
\label{tab:best_fit_and_abundance_ratios_E}
\end{table*}

\begin{table*}[ht]
\renewcommand{\arraystretch}{1.3}
\centering
\caption{Best-fit parameters from the spectral fitting of the SW region data in the 1.7-11 keV band.}
\begin{tabular}{lcccccccc}
\hline
Parameter & 1T & Log(T) & Lin(T) & Wdem & 2T Chandra & 2T EPIC & 2T & 2T Xspec \\
\hline
Si & $1.16^{+0.14}_{-0.13}$ & $1.04^{+0.12}_{-0.13}$ & $1.03^{+0.14}_{-0.12}$ & $1.09^{+0.13}_{-0.12}$ & $1.05^{+0.15}_{-0.12}$ & $1.04^{+0.12}_{-0.11}$ & $1.05^{+0.12}_{-0.13}$ & $1.00^{+0.13}_{-0.12}$ \\
S & $1.26^{+0.09}_{-0.08}$ & $1.23^{+0.08}_{-0.09}$ & $1.24^{+0.09}_{-0.09}$ & $1.26^{+0.09}_{-0.08}$ & $1.26^{+0.09}_{-0.09}$ & $1.28^{+0.08}_{-0.08}$ & $1.26^{+0.09}_{-0.08}$ & $1.21^{+0.09}_{-0.08}$ \\
Ar & $0.89^{+0.10}_{-0.09}$ & $0.92^{+0.09}_{-0.09}$ & $0.93^{+0.11}_{-0.10}$ & $0.91^{+0.10}_{-0.09}$ & $0.94^{+0.11}_{-0.10}$ & $0.93^{+0.10}_{-0.09}$ & $0.94^{+0.10}_{-0.10}$ & $0.96^{+0.10}_{-0.10}$ \\
Ca & $0.97^{+0.11}_{-0.10}$ & $1.02^{+0.11}_{-0.10}$ & $1.03^{+0.12}_{-0.11}$ & $1.00^{+0.11}_{-0.10}$ & $1.04^{+0.12}_{-0.11}$ & $1.02^{+0.10}_{-0.10}$ & $1.04^{+0.11}_{-0.11}$ & $0.94^{+0.11}_{-0.10}$ \\
Cr & $0.94^{+0.46}_{-0.42}$ & $0.93^{+0.46}_{-0.42}$ & $0.93^{+0.47}_{-0.44}$ & $0.93^{+0.46}_{-0.42}$ & $0.93^{+0.47}_{-0.45}$ & $0.91^{+0.44}_{-0.39}$ & $0.93^{+0.45}_{-0.42}$ & $0.84^{+0.44}_{-0.40}$ \\
Fe & $0.67^{+0.05}_{-0.04}$ & $0.68^{+0.04}_{-0.04}$ & $0.67^{+0.04}_{-0.04}$ & $0.67^{+0.04}_{-0.04}$ & $0.67^{+0.04}_{-0.04}$ & $0.66^{+0.04}_{-0.04}$ & $0.67^{+0.04}_{-0.04}$ & $0.69^{+0.05}_{-0.05}$ \\
Ni & $1.60^{+0.41}_{-0.38}$ & $1.53^{+0.42}_{-0.37}$ & $1.52^{+0.51}_{-0.37}$ & $1.57^{+0.40}_{-0.37}$ & $1.55^{+0.35}_{-0.44}$ & $1.48^{+0.38}_{-0.35}$ & $1.55^{+0.34}_{-0.43}$ & $1.28^{+0.40}_{-0.36}$ \\
\hline
\end{tabular}
\label{tab:best_fit_and_abundance_ratios_SW}
\end{table*}

\begin{table*}
\renewcommand{\arraystretch}{1.3}
\centering
\caption{Final Fe and X/Fe abundances in the \textit{XRISM}/Resolve pointings C, E, NW, and SW, measured with the gdem–Log(T) model.}
\begin{tabular}{lccc|ccc|ccc|ccc}
    \hline
  &  \multicolumn{3}{c}{C} & \multicolumn{3}{c}{E} & \multicolumn{3}{c}{SW} & \multicolumn{3}{c}{NW} \\
\hline
 & Val & Stat & Sys & Val & Stat & Sys & Val & Stat & Sys & Val & Stat & Sys \\
\hline
Si/Fe & 1.469 & $^{+0.135}_{-0.126}$ & $\pm0.008$ & 1.440 & $^{+0.211}_{-0.182}$ & $\pm0.056$ & 1.545 & $^{+0.196}_{-0.206}$ & $\pm0.227$ & 1.239 & $^{+0.184}_{-0.169}$ & $\pm0.031$ \\
S/Fe & 1.497 & $^{+0.115}_{-0.109}$ & $\pm0.028$ & 1.427 & $^{+0.162}_{-0.137}$ & $\pm0.054$ & 1.825 & $^{+0.164}_{-0.160}$ & $\pm0.061$ & 1.310 & $^{+0.137}_{-0.125}$ & $\pm0.016$ \\
Ar/Fe & 1.335 & $^{+0.119}_{-0.114}$ & $\pm0.035$ & 1.345 & $^{+0.186}_{-0.166}$ & $\pm0.072$ & 1.367 & $^{+0.158}_{-0.153}$ & $\pm0.098$ & 1.024 & $^{+0.141}_{-0.129}$ & $\pm0.026$ \\
Ca/Fe & 1.654 & $^{+0.149}_{-0.143}$ & $\pm0.036$ & 1.772 & $^{+0.236}_{-0.206}$ & $\pm0.106$ & 1.517 & $^{+0.187}_{-0.163}$ & $\pm0.119$ & 1.490 & $^{+0.183}_{-0.166}$ & $\pm0.038$ \\
Cr/Fe & 1.439 & $^{+0.441}_{-0.420}$ & $\pm0.033$ & 2.332 & $^{+0.766}_{-0.690}$ & $\pm0.053$ & 1.384 & $^{+0.691}_{-0.634}$ & $\pm0.013$ & 0.792 & $^{+0.512}_{-0.477}$ & $\pm0.020$ \\
Ni/Fe & 2.064 & $^{+0.482}_{-0.452}$ & $\pm0.050$ & 1.567 & $^{+0.603}_{-0.529}$ & $\pm0.041$ & 2.269 & $^{+0.629}_{-0.562}$ & $\pm0.055$ & 1.401 & $^{+0.474}_{-0.425}$ & $\pm0.017$ \\
Fe & 1.178 & $^{+0.076}_{-0.073}$ & $\pm0.026$ & 0.697 & $^{+0.060}_{-0.047}$ & $\pm0.028$ & 0.675 & $^{+0.039}_{-0.036}$ & $\pm0.040$ & 0.773 & $^{+0.057}_{-0.051}$ & $\pm0.039$ \\
\hline
\end{tabular}
\vspace{1mm}
\begin{minipage}{\textwidth}
\footnotesize
\textit{Note.} Models are fitted in the 1.7-11 keV full-FOV spectra. Uncertainties are statistical (asymmetric) and model systematic (symmetric).
\end{minipage}
\end{table*}

\subsection{Regional variations}

Here we analyze variations in abundance ratios among the four FOVs. With respect to M87, the C pointing extends out to $\sim12$~kpc, while the E, NW, and SW FOVs cover projected distances from $\sim7$~kpc to $\sim28$~kpc. Fig.~\ref{fig:Abundance_ratios_per_pointing} reveals moderate but systematic regional variations: although consistent within the errors in all four regions, the Si/Fe best-fit is the lowest in the NW; S/Fe is similar in C and E, lower in NW, and significantly enhanced in SW by 29\% on average; Ar/Fe remains constant within C, E, and SW, and is the lowest in NW by $24\%$ on average; Ca/Fe is highest in E, though consistent within the errors in all four regions; Cr/Fe is highest in E as well, consistent within 1$\sigma$ with C and SW, and lowest in the NW, although the large uncertainties limit the significance of this trend. Finally, Ni/Fe is highest in SW and C, but remains statistically consistent among all regions due to the large errors. 

Averaging over the ratios of $\alpha$-elements Si, S, Ar, Ca, over Fe, we find that the NW exhibits $\alpha$/Fe values lower by 15$\pm$4\% compared to the Center and to the East respectively, and by 19$\pm4\%$ compared to the Southwestern arm \footnote{For each region, we first compute the mean value of Si, S, Ar and Ca abundances. We then derive the corresponding uncertainty assuming that the abundances of these elements are uncorrelated. The mean abundance in each region is then normalized by the Fe abundance of the corresponding region, and the associated uncertainty on this ratio is obtained by error propagation. Finally, we compute the relative difference between the abundance ratios to Fe measured in regions C, E and SW, and that measured in the NW, expressed as a percentage relative to the value in C, E and SW respectively. The uncertainty on this relative difference is obtained by error propagation.}. So in addition to the overall enhancement of X/Fe, we observe a systematic decrease in abundance ratios from the central and arm regions toward the NW pointing. This trend is present for all measured elements, although with varying statistical significance. While the abundance ratios remain predominantly super-Solar across the full field of view, their radial decline suggests spatial variations in the chemical enrichment of the intracluster medium.

Taken together, these results demonstrate that the intracluster medium in M87 is characterized by systematically super-Solar abundance ratios, robustly detected across regions and modeling approaches. These results distinguish this system from other nearby cool-core clusters as will be discussed in Sec. \ref{sec:The Virgo cluster in the broader cluster population}, and provide a strong observational basis for a detailed discussion of the enrichment history shaping the intracluster medium in M87.

\section{Discussion}
\label{sec:Discussion}

\subsection{The Virgo cluster in the broader cluster population}
\label{sec:The Virgo cluster in the broader cluster population}

In this work, we measured the chemical abundance pattern of Si, S, Ar, Ca, Fe and Ni in the core of the Virgo cluster using \textit{XRISM}/Resolve. Its high-resolution spectroscopic capabilities enable a transformative improvement in the robustness of elemental abundance constraints compared to CCD instruments \citep[see the detailed discussion in][]{Simionescu_2019}. Observations with \textit{Hitomi}/SXS and \textit{XRISM}/Resolve have so far reported typically near-Solar abundance ratios in the probed regions of the ICM, mainly focusing on the centers of bright clusters such as Perseus, Centaurus and Abell 2029 (A2029) \citep{Hitomi_2017,Mernier_2026,Sarkar_2025}. In this context, we compare the abundance ratios measured in the Virgo cluster core with those reported for these three systems. 

\begin{figure}
    \centering
    \includegraphics[width=1\linewidth]{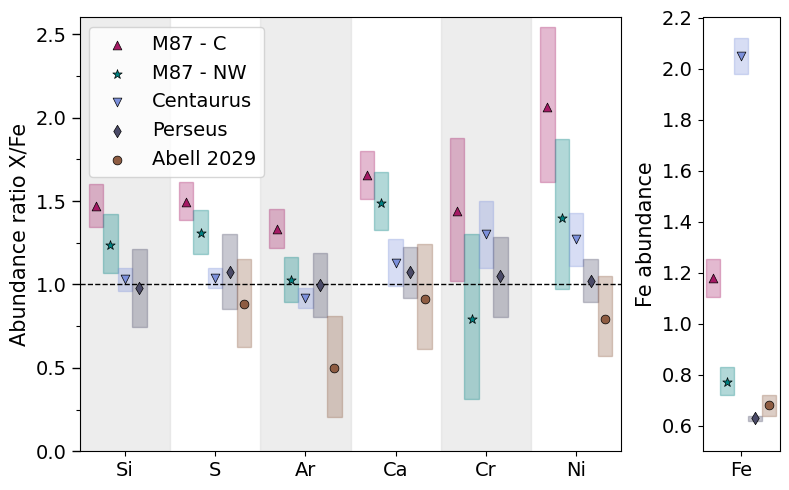}
    \caption{Comparison of elemental abundance ratios in M87 (C and NW) with those measured in A2029, Centaurus, and Perseus. Values correspond to best-fit fiducial models with 1$\sigma$ statistical uncertainties. M87 abundances are derived using the gdem–Log(T) model; Centaurus and A2029 measurements are from \textit{XRISM}/Resolve \citep{Mernier_2026,Sarkar_2025}, and Perseus from \textit{Hitomi}/SXS (model A1; bvvapec in \texttt{Xspec}, \citet{Hitomi_2017}. All abundances are given relative to the Solar reference of \citet{Lodders_2009}.}
    \label{fig:M87_Centaurus_Perseus}
\end{figure}

Fig. \ref{fig:M87_Centaurus_Perseus} presents the abundance ratios and absolute Fe abundance measured in the four clusters. For each one, we report the values identified by the original authors as the most reliable (namely Model A1 for Perseus; Table 5 for Centaurus; Table 2 for A2029). Perseus and Centaurus exhibit ratios to Fe consistent with Solar values for a wide range of elements across the tested models (Si, S, Ar, Ca, Cr, Mn, and Ni). We note that the revised Fe abundances in Perseus constrained with \textit{XRISM} data in \citet{XRISM_Perseus_2026} are consisted with \textit{Hitomi}. Similarly, A2029 shows Solar S, Ca, and Ni ratios to Fe, while Ar/Fe appears sub-Solar. Despite Centaurus hosting one of the strongest central Fe abundance peaks among nearby cool-core clusters \citep{Allen_Fabian_1994,Fukazawa_1994}, its abundance ratios remain close to Solar when averaged over the spatial scales probed by Resolve. The contrast with the super-Solar ratios observed in the Virgo core raises the question of whether M87 has an atypical chemical enrichment history, unrepresentative of the broader population of cluster cores, or whether the difference arises from other observational and/or physical effects. 

\subsection{Sensitivity of X/Fe ratios to spectral bandpass}
\label{sec:Discussion2}

\begin{figure*}
    \centering
    \includegraphics[width=1\linewidth]{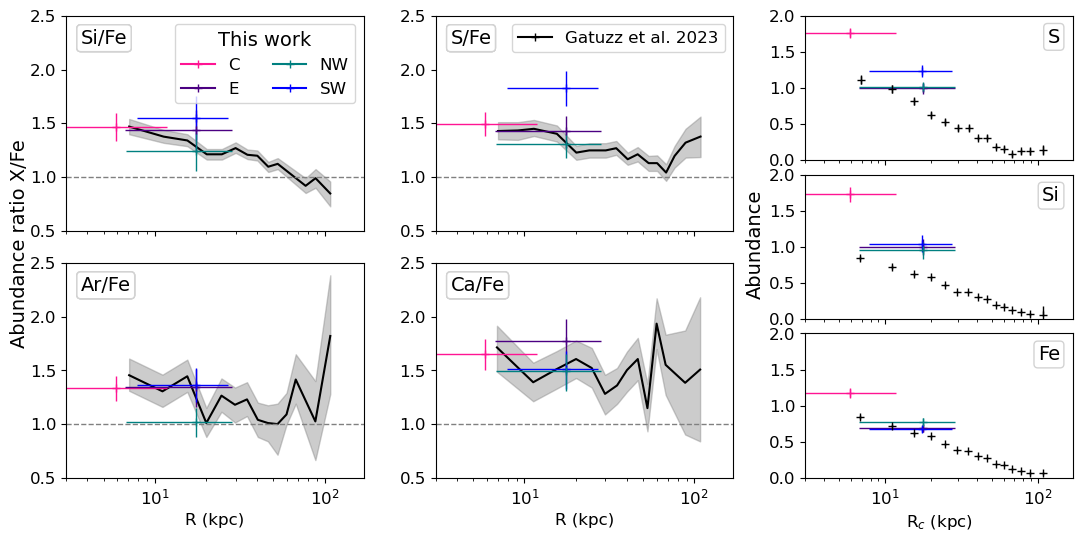}\caption{Comparison of measurements in Virgo with \textit{XRISM}/Resolve to \citet{Gatuzz_2023} with XMM-Newton/EPIC data in black (1$\sigma$ errors in gray). \textit{Left}: radial profiles of Si, S, Ar, and Ca ratios to Fe. \textit{Right}: Radial profiles of S, Si, and Fe absolute abundances.}
    \label{fig:Abundance_Gatuzz_vs_this_work}
\end{figure*}

The \textit{XRISM}/Resolve analysis is restricted to the 1.7-11 keV energy band, thus lacking diagnostics from Fe-L shell transition and low-energy lines from O, Ne and Mg below 1.7 keV. At the characteristic temperature of the Virgo cluster core ($\sim$ 1-2.5 keV), these features provide important constraints on both the temperature structure and the metal content of the ICM. This raises the question of whether the super-Solar X/Fe ratios measured by \textit{XRISM} might be biased by the absence of soft-band information. To address this, we compare the results with the recent XMM-Newton analysis of M87 by \citet{Gatuzz_2023}, who models EPIC spectra in the 0.5-4.0 keV band, thus including the missing soft X-ray diagnostics. This study employs an up-to-date atomic database and a log-normal temperature distribution to describe the multiphase ICM, and constrains the abundances of O, Ne, Si, Ar, S, Ca, Fe, and Ni relative to \citet{Lodders_2009}.

Fig.~\ref{fig:Abundance_Gatuzz_vs_this_work} compares the Si/Fe, S/Fe, Ar/Fe and Ca/Fe ratios (left panel) and the absolute radial abundance profiles of S, Si and Fe (right panel) obtained in this work with those reported by \citet{Gatuzz_2023}. Despite the limited spectral resolution of EPIC and the reduced sensitivity of the Resolve band to Fe-L emission, we find good agreement between the abundance ratios derived with both instruments. This indicates that the inferred super-Solar $\alpha$/Fe ratios are robust against differences in instrumental bandpass and spectral resolution, and are unlikely driven by the absence of Fe-L lines.

However, some differences emerge when considering absolute abundances. Specifically, Si is higher when measured with Resolve than with EPIC. The Si, S and Fe absolute abundances exhibit centrally peaked profiles, consistent with EPIC measurements, and the remaining abundances agree within uncertainties. It is not clear exactly why this difference in Si abundance occurs, but it can presumably be attributed to residual uncertainties in the XMM-Newton line spread function or effective area substructure.

\subsection{Chemical properties of the hot and cool gas phases}
\label{sec:Chemical properties hot vs. cool gas}

The core of M87 hosts a multiphase ICM with temperatures ranging from $\sim0.6$ to $\sim2$ keV \citep{Molendi_2002, Matsushita_2002, Werner_2006, Simionescu_2008, Werner_2010}. If the cooler and hotter phases possess different abundance patterns, models assuming a single metal distribution would bias the inferred X/Fe ratios. Previous studies suggest that the cool gas in M87 is spatially associated with AGN-driven uplift \citep{Churazov_2001, Million_2010, Simionescu_2026}. \citet{Young_2002} reported enhanced metal abundances in the X-ray arms observed with Chandra, interpreted as metal-rich gas buoyantly uplifted by radio bubbles. \citet{Werner_2006} detected Fe XVII emission with XMM-Newton/RGS, indicating gas as cool as $\sim$0.6-0.7 keV confined to the radio arms. \citet{Simionescu_2008} found a correlation between the fraction of gas below 1.5 keV and Fe abundance, inferring a $\sim$2.2 Solar Fe abundance of the gas in the cooler arms. These studies support a scenario in which cool, metal-rich gas from the central galaxy is uplifted along the X-ray bright structures mapped by the E and SW FOVs. 

In this work, the center shows systematically super-Solar abundance ratios, while the NW exhibits lower ratios representative of the ambient ICM less affected by AGN-driven perturbations. The E and SW arms display intermediate values, closer to the center than to the NW, suggesting azimuthal asymmetries driven by AGN uplift and incomplete mixing with the ambient cluster atmosphere. We therefore test whether chemically distinct gas phases can explain these variations by fitting the E and SW with a multi-temperature, multi-abundance model that we name 2T-2Z. Assuming the cooler gas originates from M87, i.e. shares the same abundance pattern measured in the C pointing, we fix its abundances of Si, S, Ar, Ca, Cr, Fe and Ni to the central values while allowing its EM to vary. For the hotter component of this 2T-2Z model, abundances of Si, S, Ar, Ca, Cr, Fe and Ni and EM are free to vary. Both CIEs have their temperature and velocity free to vary as well. 

\begin{figure}
    \centering
    \includegraphics[width=1\linewidth]{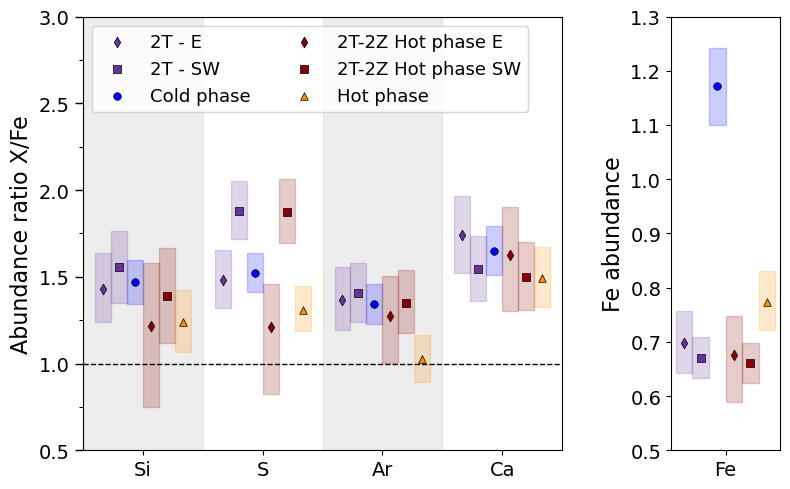}
    \caption{\textit{Left:} Abundance ratios of Si, S, Ar, and Ca relative to Fe. \textit{Right}: Absolute Fe abundance. Values measured in six cases. 2T - E and SW: ratios obtained with the 2T model in the E and SW regions, assuming both gas phases share the same chemical composition. Cold phase: ratios measured in region C of M87 with the 2T model. 2T-2Z metallicity of the cooler phase is fixed to these values. 2T-2Z Hot phase in E and SW: ratios measured in the hotter CIE component with the 2T-2Z model. Hot phase: ratios in region NW obtained with the gdem-Log(T) model.}
    \label{fig:2T 2Z plot}
\end{figure}

The results are shown in Fig.~\ref{fig:2T 2Z plot}. With a two-temperature model assuming a single abundance pattern (2T), the E and SW abundance ratios differ from the NW ("hot phase"): the average $\alpha$/Fe ratio in the east is higher by $\sim19\pm11\%$, and in the southwest by $\sim26\pm11\%$ relative to the NW. When the cooler-phase metallicity is fixed to the central pattern, the hotter-phase ratios in E and SW shift toward the NW values. The hot-phase metallicity in E is then higher by only $\sim5\pm14\%$, and in SW by $\sim21\pm11\%$, compared to the NW. This behavior suggests that the arms contain chemically distinct cool gas embedded in a hotter ICM with an abundance pattern closer to the relaxed cluster atmosphere. The weaker drop observed in the SW may reflect its smaller hot-to-cold gas fraction compared to the eastern arm.

The observed abundance diversity can be explained by AGN-driven uplift of cool and metal-rich gas from the galaxy center, or by incomplete mixing between uplifted gas and ambient ICM. These results support the scenario in which AGN-driven uplift transports centrally enriched gas along the X-ray bright arms, where it remains only partially mixed with the surrounding ICM. Indeed, once this chemically distinct cool gas phase is accounted for, the hot component appears more chemically homogeneous across azimuth, though it still remains super-Solar. Overall, accounting explicitly for multi-abundance structures slightly reduces the azimuthal discrepancies and suggests that the large-scale hot ICM is closer to chemically uniform, and closer to Solar values, than implied by single-metallicity, multi-temperature modeling. This demonstrates that part of the apparent abundance diversity in M87 arises from the potential coexistence of chemically distinct gas phases. 

\subsection{Sensitivity of X/Fe ratios to probed spatial scales}

As shown in the previous sections, the super-Solar $\alpha$/Fe ratios measured in the central regions of M87 differ from the near-Solar values reported for A2029, Centaurus and Perseus. Extended to the most recent results at CCD resolution, those near-Solar patterns are in line with virtually all investigated nearby cool-core clusters so far \citep[e.g.][]{Simionescu_2009,Mernier_2017}, making the case of Virgo truly unique. Since Virgo is also the nearest cluster to us ($z=0.0043$; compared to Centaurus, Perseus, and A2029 with redshifts of $0.0104$, $0.017$, and $0.0767$, respectively), a question arises as to whether this different behavior could be naturally explained by our access to substantially smaller physical scales in the Virgo core. In fact, while the $\sim3'$ Resolve FOV corresponds approximately to 36~kpc, 66~kpc, and 267~kpc in Centaurus, Perseus, and A2029 respectively, the same angular size pointed towards Virgo corresponds corresponds to merely $\sim15$~kpc. 

As discussed further in Sect.~\ref{sec:discussion:history}, there may well be a profound connection between the stellar composition of clusters' BCG and the ICM composition of regions pervading that BCG. In the case of M87, \citet{Gu_2022} report O/Fe and Mg/Fe stellar ratios of more than twice the Solar value (about 2.4 and 2.5 Solar, respectively; see also \citealt{Sarzi_2018,Parikh_2024}). Assuming consistent values for heavier $\alpha$/Fe ratios such as Si/Fe and S/Fe and a mixture of X-ray gas coming from (Solar) pre-enriched ICM and stellar mass loss from M87 itself, it is plausible to recover the $\sim$1.5~Solar values seen in the ratios observed by Resolve.

Whereas this possibility could in principle explain our super-Solar $\alpha$/Fe ratios, we note that the stellar half-light radius of Centaurus' BCG, NGC4696, is about $35$~kpc \citep{ArnalteMur_2006}, i.e. $\sim$2.7~arcmin. Although this scale is clearly reachable by Resolve, no significant deviation from a Solar composition is reported through detailed sub-array analysis (Fukushima et al. in prep). Similarly, \citet{Sani_2018} report a half-light radius of Perseus' BCG, NGC1275, of $45 \pm 15$~kpc, thus extending to about 1.3--2.7~arcmin on the sky. Albeit subject to more uncertainties (due to the bright X-ray emission of its central AGN), no super-Solar pattern is observed with \textit{XRISM} across such scales either \citep{XRISM_Centaurus_2026}.

\subsection{Enrichment history}\label{sec:discussion:history}

Besides the potential multi-metallicity structure of the gas and the smaller physical spatial scales probed in M87, we must also consider whether a different enrichment history may be responsible for the observed super-Solar $\alpha$/Fe ratios in M87, as opposed to most other targets showing Solar abundance patterns.

A detailed multi-wavelength view of M87, NGC1275 (the BCG of Perseus), and NGC4696 (the BCG of Centaurus), indeed reveals clear differences between these galaxies. For example, with an estimated $H_2$ mass of only $4.7\times10^5 M_\odot$ \citep{Simionescu_2018_ALMA}, M87 has a cold molecular gas reservoir orders of magnitude smaller than NGC4696 ($\sim10^8 M_\odot$, \citealt{Olivares_2019}) and NGC1275 ($4\times10^{10} M_\odot$, \citealt{Salome_2006}), while in A2029's BCG only an upper limit of $\lesssim1.7\times10^{9} M_\odot$ was found \citep{Salome_Combes_2003}. Hubble Space Telescope images of these BCGs further reveal recent star formation in the Perseus Cluster core \citep{Canning_2010}, and spectacular dust lanes in Centaurus \citep{Fabian_2016}; neither of these are present in M87, which shows a regular "red and dead" morphology. It is therefore natural to conclude that the recent star formation history of these three BCGs is rather different, hence leading to different enrichment patterns also in their hotter gas atmospheres. At first glance, one may naively expect those BCGs that have a larger mass of cold gas to have enhanced $\alpha$/Fe ratios. This is because, when more molecular gas is available to form young stars, the most massive of these should explode within a short time after the starburst as CC SNe, which are a known source of $\alpha$ elements, and contribute only very little iron (see \citet{Nomoto_2013} for a review). In terms of the ICM abundances, we observe the opposite, with the cold gas-poorest target, M87, having instead the highest $\alpha$/Fe ratios. 

A perhaps plausible alternative scenario would be to consider that the typical enrichment of the central ICM surrounding the BCG is rather driven by stellar winds, and that the products from recently exploded SNe do not quickly mix into the hotter atmosphere, but instead remain locked in the cooler ISM of each galaxy, and are not observed at X-ray wavelengths. In terms of the stellar abundances, it is known that short, concentrated bursts of star formation happening early in the history of the galaxy lead to higher $\alpha$/Fe ratios, while a longer star formation period brings stellar compositions (and hence that of their stellar winds) closer to the Solar pattern (see discussion in e.g. \citet{Conroy_2014}). In this scenario, BCGs that are still actively forming stars, like NGC1275, would lead to central ICM enrichment peaks that are closer to Solar composition, while older populations like M87 would correspond to higher $\alpha$/Fe in the surrounding ICM as reported here.

Before drawing firm conclusions on the favored enrichment model, we must however caution the reader that the number of clusters for which microcalorimeter-precision abundance patterns are available so far is still very limited. Moreover, \textit{XRISM} observations of A2029 still show Solar $\alpha$/Fe ratios, although like M87, this target also lacks cold gas, with only an upper limit on molecular emission reported by \citet{Salome_Combes_2003}. In addition to enhanced $\alpha$/Fe ratios, this work's results also show super Solar Ni/Fe, consistent with XMM-Newton/RGS data in \citet{Fukushima_2023}. Given that both Fe and Ni come almost exclusively from type Ia SNe, and that Ni/Fe was reported to be Solar in the three other clusters, this could support a biased measurement mostly of the Fe abundance rather than genuinely high $\alpha$/Fe ratios. A growing sample of \textit{XRISM} abundance pattern measurements, including both spatial mapping in nearby systems, and a larger number of targets, will be essential to elucidate the underlying reason for any diversity in chemical composition of the central ICM.

\section{Conclusion}
\label{sec:Conclusion}

We have investigated the chemical composition in four regions of the Virgo cluster core with \textit{XRISM}/Resolve. The high spectral resolution of Resolve enables precise measurements of elemental abundance ratios, allowing us to investigate regional variations and to assess systematic effects related to multiphase structure and bandpass limitations. We summarize our main results: 
\begin{itemize}
    \item The central, eastern and southwestern pointings exhibit systematically super-Solar ratios of Si, S, Ar, Ca, Cr and Ni to Fe, in contrast to the Solar abundance patterns reported for A2029, Centaurus and Perseus. This establishes Virgo as chemically distinct among this small sample of clusters observed with high-resolution spectroscopy.
    \item The northwest region, which is not affected by AGN activity, displays lower ratios, closer to the Solar values, and traces the more relaxed, cluster-wide ICM. 
    \item Multi-temperature, multi-abundance modeling suggests that part of the X-ray emission from the arms arises from cool gas with a central-like composition embedded in a hotter phase whose abundance pattern is closer to that of the northwest. This behavior is consistent with AGN-driven uplift of chemically distinct gas and incomplete mixing with the ambient ICM.
    \item Because the Resolve analysis is restricted to the 1.7–11 keV band, it lacks diagnostics of Fe-L and other soft-band lines. However, comparison with XMM-Newton/EPIC measurements covering 0.5–4 keV shows good agreement in terms of abundance ratios. This indicates that the super-Solar pattern in M87 is not driven by temperature biases associated with the restricted bandpass.
    \item The proximity of Virgo allows \textit{XRISM} to probe smaller physical scales, comparable to the stellar extent of M87, potentially revealing localized enrichment signatures. Nevertheless, sub-field analyses of Centaurus probing similar scales do not show enhanced X/Fe ratios. Based on this limited sample, spatial resolution effects might not fully account for the chemical differences between Virgo and other clusters.
    \item The distinct abundance pattern in M87 may reflect a different enrichment history of its BCG compared to systems such as Centaurus and Perseus. In particular, M87’s old stellar population and lack of significant molecular gas suggest a star formation history dominated by early, short-lived bursts, which are expected to produce $\alpha$-enhanced stellar populations. If stellar mass loss plays a dominant role in enriching the central ICM, the hot gas may inherit this $\alpha$-enhanced signature. However, the limited number of clusters with microcalorimeter-quality abundance measurements cautions against over-interpretation at this stage. 
\end{itemize}
Overall, our results demonstrate that the chemical composition of cluster cores can vary significantly on small spatial scales and may differ from the near-Solar pattern observed in several well-studied systems. The Virgo cluster highlights the importance of spatially resolved, high-resolution X-ray spectroscopy to disentangle the roles of stellar enrichment, AGN-driven transport, multiphase structure, and mixing in shaping the metal distribution of the ICM. Expanding the sample of clusters observed with \textit{XRISM}, and in the future with NewAthena, will be crucial to determine whether Virgo represents an outlier or reveals a broader diversity in the enrichment histories of cluster cores.

\section*{Data Availability}
The XRISM data analyzed in this work are publicly available from the JAXA DARTS archive (https://data.darts.isas.jaxa.jp/pub/xrism/data/obs/) and the NASA HEASARC XRISM archive.

\begin{acknowledgements}
This work is based on observations obtained with \textit{XRISM}, XMM-Newton, and Chandra. The \textit{XRISM} mission is a JAXA/NASA/ESA collaboration and we thank the \textit{XRISM} Science Team for their support. This work was supported by JSPS Core-to-Core Program (grant number:JPJSCCA20220002). Contributions from M. Loewenstein are based upon work supported by NASA under award number 80GSFC24M0006. The material is based upon work supported by NASA under award number 80GSFC21M0002. XMM-Newton is an ESA science mission with instruments and contributions directly funded by ESA Member States and NASA. Chandra data were obtained from the Chandra X-ray Observatory and its Data Archive. We acknowledge the use of the \texttt{SPEX} spectral fitting package and the \texttt{SPEXACT} atomic database. The \texttt{SPEX} team is thanked for maintaining the software and providing support for high-resolution X-ray spectroscopy analysis. SRON is supported financially by The Netherlands Organisation for Scientific Research (Nederlandse Organisatie voor Wetenschappelijk Onderzoek, NWO). We thank the referee for their useful comments which contributed to increase the overall quality and readiness of this manuscript. We also acknowledge the use of Python packages matplotlib and numpy for data analysis and visualization. 

\end{acknowledgements}

\bibliographystyle{aa}
\bibliography{references.bib}

\begin{appendix}

\section{Systematic tests}
\label{sec:Appendix_A_systematic_tests}

We have explored the impact of two key factors on the measured X/Fe ratios: the assumptions regarding the thermal structure of the ICM (Sect.~\ref{sec:Impact of temperature structure}) and the choice of atomic database (Sect.~\ref{sec:Impact of atomic database}). As discussed in Sect.~\ref{sec:Impact of atomic database}, the effect of using \texttt{SPEXACT} (\texttt{SPEX}) vs. AtomDB (\texttt{Xspec}) on X/Fe values is comparable in magnitude to the systematics introduced by different temperature-structure models. Systematic tests can, in principle, be performed equivalently in \texttt{SPEX} or \texttt{Xspec}; however, \texttt{Xspec} offers a more straightforward workflow. Indeed, in \texttt{SPEX}, each test requires reconverting the data to \texttt{SPEX} format, substantially increasing computation time. Moreover, thanks to the analysis presented in \citetalias{Simionescu_2026}, all relevant models for the systematics tests have already been run in \texttt{Xspec}. For these reasons, we adopt their \texttt{Xspec} analysis as our reference framework. Using this setup, we quantify statistical uncertainties arising from PSF contamination from the central pointing, non X-ray background (NXB) level, and instrumental calibration through ARF adjustments.

The PSF contamination from the bright central pointing, including emission from the AGN/jet and unresolved LMXBs, contaminates the spectra of the outer regions. To quantify this effect, we varied the normalizations of the PSF contamination from region C by $\pm 30\%$ around their best-fit values and re-derived the abundance ratios. The resulting variations in the abundance ratios remain within the combined statistical and systematic uncertainties, indicating that PSF contamination does not dominate the error budget.

The NXB contributes additional counts to the spectra, particularly at energies above $\sim$8 keV in in regions of lower cluster surface brightness (E, SW, and NW). To quantify this effect, the NXB normalization was varied by $\pm 10\%$ around its nominal value and X/Fe ratios re-derived. The resulting variations in the abundance ratios negligible compared to both statistical and other systematic uncertainties, confirming that NXB uncertainties are not a limiting factor for this analysis.

We explored the effect of the systematic uncertainties of the Resolve effective area on the abundance ratio measurements. We applied the modification factor to Resolve spectra \citep{XRISM2025_NGC3783} that is used to bring the Resolve effective area into agreement with CCDs to account for the soft band ($\leq~3keV$) excess seen in the NGC 3783 spectra (about 20\% higher than Xtend). Since the soft energy effective area calibration is ongoing, \citet{XRISM2025_NGC3783} provides the most up-to-date method. This systematic error would mainly affect the Si/Fe, S/Fe, Ar/Fe, and Ca/Fe ratios, where the correction factor alters the effective area by 13\% at the line energy of Si, 5\% for S, and less than 2\% for Ar and Ca. At higher energies where Cr/Fe and Ni/Fe are concerned, the effective area systematic uncertainty for Cr, Ni and Fe are less than 2\% \citep{Hayashi_XMA}. We note that, even after accommodating a generous correction factor, Solar abundances are still recovered. The resulting variations in abundance ratios are modest, and do not alter our main conclusions.

\begin{figure}
    \centering
    \includegraphics[width=1\linewidth]{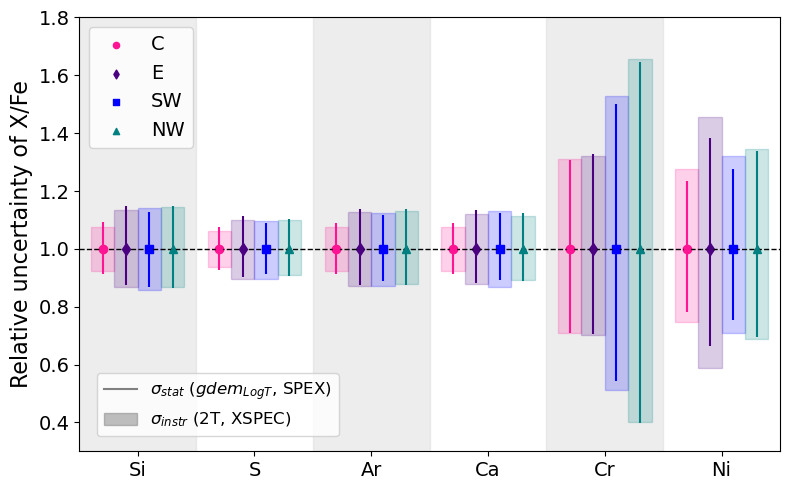}
    \caption{Relative uncertainties on the elemental abundance ratios (X/Fe) measured in the Virgo cluster with XRISM/Resolve in the 1.7-11 keV band. The data are normalized to the central best-fit values for each element, so that a value of 1 corresponds to no deviation. Markers show the statistical uncertainties derived from the fiducial \texttt{gdem Log(T)} model in \texttt{SPEX}, while the rectangles indicate the instrumental systematics derived from \texttt{Xspec} (based on the 2T model in C, E and SW, and 1T in NW).}
    \label{fig:Abundance_ratios_SPEX_vs_Xspec}
\end{figure}

We plot the combined instrumental systematics from these tests --PSF contamination from the central pointing, non NXB level, and instrumental calibration through ARF adjustments-- in quadrature and compare them to the statistical error on the fiducial model on Fig.~\ref{fig:Abundance_ratios_SPEX_vs_Xspec}. The plot shows that these two sources of errors are comparable in magnitude. Therefore, the abundance ratios measured from XRISM/Resolve are robust with respect to assumptions about the ICM temperature structure, but limited by the combination of instrumental and statistical uncertainties.

\section{CLUS model parameters}

Spectra for the CLUS model \citep{Stefanova_2025} are computed by projecting a spherically symmetric model of a galaxy cluster onto the sky. The cluster model specifies the radial profiles of the temperature, T(r), density, n(r), and metallicity, Z(r), where the radius r is measured from the cluster center. The model cluster is divided into spherical shells, where the X-ray emission is computed using a collisional ionisation equiibrium model for the plasma. To compute the spectrum of a given region on the sky, the contributions from each of the overlapping shells are then summed. Table~\ref{CLUS parameters} gives the initial parameters of the model. All parameters not explicitly listed here are set to their respective model defaults.

\begin{table}[h]
\centering
\caption{CLUS model initial parameters for the XRISM/Resolve NW pointing in the Virgo cluster.}
\label{CLUS parameters}
\begin{tabular}{cc|cc}
\hline
Parameter & Value                         & Parameter & Value \\ \hline
$n_{1}$   & $310800 \text{ m}^{-3}$       & $ct$      & $10.0$ \\
$r_{tc1}$ & $2.06 \times 10^{-3} r_{500}$ & $A$       & $1.48$ \\
$\beta_1$ & $2.0$                         & $B$       & $0.02 r_{500}$ \\
$n_{2}$\textsuperscript{*} & $92783 \text{ m}^{-3}$        & $C$       & $0.72$ \\
$r_{tc2}$ & $4.99 \times 10^{-3} r_{500}$ & $D$       & $0.16$ \\
$\beta_2$ & $0.431$                       & $E$       & $1.00 \times 10^{-7} r_{500}$ \\
$T_h$     & $2.61 \text{ keV}$             & $F$       & $0.49 r_{500}$ \\
$T_c$\textsuperscript{\textdagger}& $0.1 \text{ keV}$ & $G$    & $0.15$ \\
$r_{to}$  & $8.9 r_{500}$                & $r_{500}$ & $2.045 (10^{22} \text{ cm})^{(a)}$ \\
$r_{tc}$  & $0.07 r_{500}$               & $r_{out}$ & $1$ \\
$\mu$     & $0.52$                       & $r_{min}$ & $3.5 \times 10^{-3}$ \\
$at$      & $0.12$                       & $r_{max}$ & $0.02$ \\
$bt$      & $1.4$                         \\
\hline
\multicolumn{4}{l}{\small \textsuperscript{*} Note for $n_{2}$: In the fit, $n_{2}$ is coupled to $n_{1}$ by a factor of $n_{2}/n_{1}=0.337$} \\
\multicolumn{4}{l}{\small \textsuperscript{\textdagger} Note for $T_c$: In the fit, $T_c$ is coupled to $T_h$ by a factor of $T_c/T_h=0.04$} \\
\end{tabular}
\end{table}

\begin{figure*}
    \centering
    \includegraphics[width=1\linewidth]{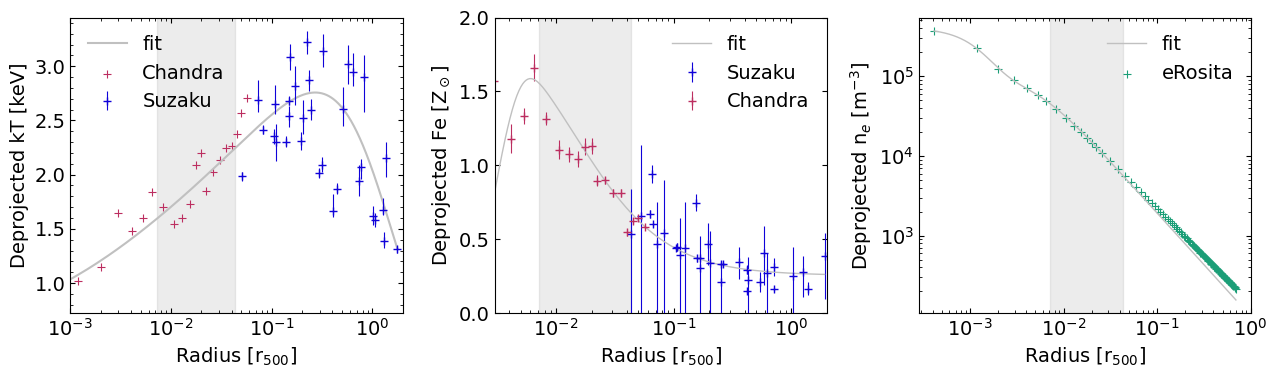}
    \caption{Density profiles used in the CLUS model definition. The gray region on each subplot highlights the spatial extent of the NW \textit{XRISM}/Resolve pointing, where the CLUS model has been tested. Suzaku temperature and Fe abundance profiles from \citet{Simionescu_2017_Suzaku}. Temperature profile from Chandra data and Fe abundance profiles from eROSITA density profile from \citetalias{XRISM_2025}. }
    \label{fig:CLUS_profiles}
\end{figure*}

\section{Best-fit parameters}
\label{Best fit parameters}

We report in the following table the best fit parameters as obtained with the various modeling strategies presented in \ref{sec:Spectral modeling}.

\begin{table*}[ht]
\renewcommand{\arraystretch}{1.2}
\centering
\begin{tabular}{lcccccc}
\hline
Parameter & 1T & 1T Xspec & Log(T) & Lin(T) & Wdem & CLUS \\
\hline
$kT, T_h$ & $2.20^{+0.10}_{-0.01}$ & $2.13^{+0.07}_{-0.05}$ & $2.20^{+0.10}_{-0.01}$ & $2.20^{+0.10}_{-0.01}$ & $2.50^{+0.20}_{-0.20}$ & $4.80^{+0.10}_{-0.10}$ \\
$\sigma_T$ & - & - & $0.10^{+0.01}_{-0.01}$ & $0.60^{+0.20}_{-0.10}$ & - & - \\
$z$ & $4.22^{+0.02}_{-0.02}$ & $4.21^{+0.02}_{-0.02}$ & $4.22^{+0.02}_{-0.02}$ & $4.22^{+0.02}_{-0.02}$ & $4.22^{+0.02}_{-0.02}$ & $4.22^{+0.02}_{-0.02}$ \\  
$v_{rms}$ & $62.79^{+13.09}_{-14.31}$ & $29.55^{+24.55}_{-29.55}$ & $65.53^{+13.03}_{-13.96}$ & $65.00^{+12.96}_{-14.04}$ & $59.38^{+13.72}_{-15.79}$ & $89.13^{+18.37}_{-20.38}$ \\
$log_{10}(LMXB)$ & $-12.48^{+0.11}_{-0.01}$ & $-12.30^{+0.09}_{-0.15}$ & $-12.59^{+0.15}_{-0.01}$ & $-12.57^{+0.15}_{-0.01}$ & $-12.53^{+0.13}_{-0.01}$ & $-12.57^{+0.13}_{-0.01}$ \\
\hline
Cstat & $4565.7$ & $3605.41$ & $4563.7$ & $4563.7$ & $4564.3$ & $4564.6$ \\
d.o.f. & 4390 & 3434 & 4389 & 4389 & 4389 & 4391 \\
\hline
\end{tabular}
\caption{Best-fit parameters from the spectral fitting of the NW region data in the 1.7-11 keV band. LMXB flux in 2-7 keV band.}
\end{table*}

\begin{table*}[ht]
\renewcommand{\arraystretch}{1.2}
\centering
\begin{tabular}{lcccccccc}
\hline
Parameter & 1T & Log(T) & Lin(T) & Wdem & 2T Chandra & 2T EPIC & 2T & 2T Xspec \\
\hline
$kT$ & $1.80^{+0.01}_{-0.01}$ & $1.70^{+0.01}_{-0.01}$ & $1.70^{+0.01}_{-0.01}$ & $2.20^{+0.10}_{-0.10}$ & $1.93^{+0.03}_{-0.03}$ & $1.83^{+0.02}_{-0.02}$ & $1.97^{+0.15}_{-0.07}$ & $1.97^{+0.17}_{-0.07}$ \\
$kT_{cold}$ & - & - & - & - & $1.24^{+0.01}_{-0.01}$ & $1.11^{+0.01}_{-0.01}$ & $1.23^{+0.19}_{-0.20}$ & $1.30^{+0.17}_{-0.07}$ \\
$\sigma_T$ & - & $0.20^{+0.01}_{-0.01}$ & $0.90^{+0.10}_{-0.10}$ & - & - & - & - & - \\
$z$ & $4.27^{+0.02}_{-0.02}$ & $4.28^{+0.02}_{-0.02}$ & $4.28^{+0.02}_{-0.02}$ & $4.28^{+0.02}_{-0.02}$ & $4.36^{+0.03}_{-0.03}$ & $4.32^{+0.02}_{-0.02}$ & $4.37^{+0.07}_{-0.04}$ & $4.43^{+0.10}_{-0.04}$ \\
$z_{cold}$ & - & - & - & - & $3.86^{+0.12}_{-0.12}$ & $3.54^{+0.24}_{-0.22}$ & $3.93^{+0.15}_{-0.16}$ & $3.84^{+0.24}_{-0.14}$ \\
$v_{rms}$ & $148.00^{+8.91}_{-8.97}$ & $150.21^{+8.68}_{-8.86}$ & $150.15^{+8.89}_{-8.77}$ & $145.73^{+8.94}_{-9.14}$ & $141.01^{+9.63}_{-10.46}$ & $139.71^{+10.14}_{-10.10}$ & $142.33^{+9.67}_{-9.71}$ & $150.71^{+13.69}_{-24.82}$ \\
$v_{rms, cold}$ & - & - & - & - & - & - & - & $153.26^{+40.12}_{-46.69}$ \\
$log_{10}(LMXB)$ & $-11.86^{+0.04}_{-0.01}$ & $-11.93^{+0.05}_{-0.01}$ & $-11.92^{+0.06}_{-0.01}$ & $-11.92^{+0.05}_{-0.01}$ & $-11.91^{+0.04}_{-0.01}$ & $-11.88^{+0.04}_{-0.01}$ & $-11.92^{+0.04}_{-0.01}$ & $-11.81^{+0.05}_{-0.05}$ \\
\hline
Cstat & $2511.5$ & $2496.4$ & $2494.7$ & $2491.6$ & $2483.5$ & $2490.5$ & $2482.3$ & $1921.6$ \\
d.o.f. & 2444 & 2443 & 2443 & 2442 & 2443 & 2443 & 2441 & 1986 \\
\hline
\end{tabular}
\caption{Best-fit parameters from the spectral fitting of the C region data in the 1.7-11 keV band. LMXB flux in 2-7 keV band.}
\end{table*}

\begin{table*}[ht]
\renewcommand{\arraystretch}{1.2}
\centering
\begin{tabular}{lcccccccc}
\hline
Parameter & 1T & Log(T) & Lin(T) & Wdem & 2T Chandra & 2T EPIC & 2T & 2T Xspec \\
\hline
$kT$ & $2.00^{+0.01}_{-0.01}$ & $1.80^{+0.10}_{-0.10}$ & $1.90^{+0.10}_{-0.10}$ & $2.94^{+0.20}_{-0.17}$ & $2.28^{+0.07}_{-0.07}$ & $2.07^{+0.05}_{-0.05}$ & $2.38^{+0.26}_{-0.14}$ & $2.60^{+0.52}_{-0.23}$ \\
$kT_{cold}$ & - & - & - & - & $1.31^{+0.02}_{-0.02}$ & $1.13^{+0.01}_{-0.01}$ & $1.20^{+0.24}_{-0.18}$ & $1.60^{+0.11}_{-0.11}$ \\
$\sigma_T$ & - & $0.20^{+0.01}_{-0.01}$ & $1.10^{+0.10}_{-0.10}$ & - & - & - & - & - \\
$z$ & $4.30^{+0.03}_{-0.03}$ & $4.31^{+0.03}_{-0.03}$ & $4.31^{+0.03}_{-0.03}$ & $4.31^{+0.03}_{-0.03}$ & $4.28^{+0.04}_{-0.04}$ & $4.28^{+0.04}_{-0.03}$ & $4.29^{+0.04}_{-0.04}$ & $4.23^{+0.04}_{-0.04}$ \\
$z_{cold}$ & - & - & - & - & $4.46^{+0.14}_{-0.14}$ & $4.70^{+0.39}_{-0.39}$ & $4.40^{+0.12}_{-0.12}$ & $4.49^{+0.10}_{-0.08}$ \\
$v_{rms}$ & $78.90^{+15.90}_{-15.77}$ & $89.33^{+15.96}_{-14.73}$ & $88.73^{+16.58}_{-15.05}$ & $76.24^{+16.80}_{-18.23}$ & $78.05^{+17.14}_{-17.58}$ & $75.54^{+17.21}_{-17.80}$ & $82.00^{+16.90}_{-17.04}$ & $29.55^{+24.55}_{-29.55}$ \\
$v_{rms, cold}$ & - & - & - & - & - & - & - & $160.26^{+27.52}_{-31.08}$ \\
$log_{10}(LMXB)$ & $-12.49^{+0.10}_{-0.01}$ & $-13.56^{+0.72}_{-0.01}$ & $-13.19^{+0.38}_{-0.01}$ & $-13.20^{+0.44}_{-0.01}$ & $-12.87^{+0.24}_{-0.01}$ & $-12.61^{+0.13}_{-0.01}$ & $-13.06^{+0.35}_{-0.01}$ & $-12.54^{+0.16}_{-0.31}$ \\
\hline
Cstat & $2417.3$ & $2388.0$ & $2386.8$ & $2386.4$ & $2390.8$ & $2401.5$ & $2386.9$ & $1852.0$ \\
d.o.f. & 2319 & 2318 & 2318 & 2317 & 2318 & 2318 & 2317 & 1884 \\
\hline
\end{tabular}
\caption{Best-fit parameters from the spectral fitting of the E region data in the 1.7-11 keV band. LMXB flux in 2-7 keV.}
\end{table*}

\begin{table*}[ht]
\renewcommand{\arraystretch}{1.2}
\centering
\begin{tabular}{lcccccccc}
\hline
Parameter & 1T & Log(T) & Lin(T) & Wdem & 2T Chandra & 2T EPIC & 2T & 2T Xspec \\
\hline
$kT$ & $2.20^{+0.01}_{-0.01}$ & $2.10^{+0.01}_{-0.01}$ & $2.10^{+0.01}_{-0.01}$ & $2.60^{+0.10}_{-0.10}$ & $2.35^{+0.04}_{-0.04}$ & $2.29^{+0.04}_{-0.04}$ & $2.35^{+0.08}_{-0.06}$ & $2.47^{+0.28}_{-0.20}$ \\
$kT_{cold}$ & - & - & - & - & $1.31^{+0.03}_{-0.03}$ & $1.06^{+0.01}_{-0.01}$ & $1.32^{+0.18}_{-0.20}$ & $1.60^{+0.11}_{-0.11}$ \\
$\sigma_T$ & - & $0.10^{+0.01}_{-0.01}$ & $0.60^{+0.10}_{-0.20}$ & - & - & - & - & - \\
$z$ & $4.08^{+0.03}_{-0.03}$ & $4.08^{+0.03}_{-0.03}$ & $4.08^{+0.03}_{-0.03}$ & $4.08^{+0.03}_{-0.03}$ & $4.15^{+0.03}_{-0.03}$ & $4.12^{+0.03}_{-0.03}$ & $4.15^{+0.04}_{-0.04}$ & $4.20^{+0.04}_{-0.04}$ \\
$z_{cold}$ & - & - & - & - & $3.46^{+0.18}_{-0.16}$ & $3.34^{+0.18}_{-0.18}$ & $3.46^{+0.20}_{-0.22}$ & $3.65^{+0.21}_{-0.32}$ \\
$v_{rms}$ & $83.19^{+14.89}_{-15.66}$ & $86.04^{+14.45}_{-14.78}$ & $85.80^{+15.08}_{-15.35}$ & $79.74^{+15.59}_{-16.71}$ & $70.96^{+17.76}_{-19.36}$ & $75.84^{+14.71}_{-15.77}$ & $70.96^{+15.74}_{-21.47}$ & $29.55^{+24.55}_{-29.55}$ \\
$v_{rms, cold}$ & - & - & - & - & - & - & - & $160.26^{+27.52}_{-31.08}$ \\
$log_{10}(LMXB)$ & $<-12.87$ & $<-12.88$ & $<-12.88$ & $<-13.00$ & $<-13.10$ & $<-13.07$ & $<-12.88$ & $<-12.93$ \\
\hline
Cstat & $4544.2$ & $4538.5$ & $4538.3$ & $4538.2$ & $4537.1$ & $4567.2$ & $4537.0$ & $1853.5$ \\
d.o.f. & 4372 & 4370 & 4370 & 4369 & 4371 & 4371 & 4369 & 1701 \\
\hline
\end{tabular}
\caption{Best-fit parameters from the spectral fitting of the SW region data in the 1.7-11 keV band. 2$\sigma$ upper limits LMXB flux.}
\end{table*}

\end{appendix}
\end{document}